\documentclass[twocolumn, trackchanges]{aastex63}

\usepackage[utf8]{inputenc}
\usepackage{multirow}
\usepackage{booktabs}

\newcommand{\Mjup}{\,\ensuremath{\mathrm{M}_\mathrm{Jup}}\,}
\newcommand{\Rjup}{\,\ensuremath{\mathrm{R}_\mathrm{Jup}}\,}
\newcommand{\Msun}{\,\ensuremath{\mathrm{M}_{\odot}}\,}
\newcommand{\Rsun}{\,\ensuremath{\mathrm{R}_{\odot}}\,}
\newcommand{\mps}{\ensuremath{\text{m s}^{-1}}\,}

\newcommand{\teff}{$T_{\rm eff}$}
\newcommand{\logg}{$\log g$}
\newcommand{\vsini}{$v\sin i$}
\newcommand{\feh}{[Fe/H]}
\newcommand{\tna}{$^*$}
\newcommand{\tnb}{$^\dag$}
\newcommand{\tnc}{$^\ddag$}
\newcommand{\tnd}{$^\S$}
\newcommand{\tne}{$^\times$}
\newcommand{\tnf}{$^\parallel$}
\newcommand{\kms}{\hbox{km\,s$^{-1}$}}
\shorttitle{TOI--1278\,B}
\shortauthors{Artigau et al.}

\begin{document}

\title{TOI--1278\,B: SPIRou unveils a rare Brown Dwarf Companion in Close-In Orbit around an M dwarf}
\correspondingauthor{\'Etienne Artigau}
\email{etienne.artigau@umontreal.ca}

\author[0000-0003-3506-5667]{\'Etienne Artigau}% CONFIRMATION EMAIL
\affiliation{Universit\'e de Montr\'eal, D\'epartement de Physique, IREX, Montr\'eal, QC H3C 3J7, Canada}
\affiliation{Observatoire du Mont-M\'egantic, Universit\'e de Montr\'eal, Montr\'eal H3C 3J7, Canada}

\author[0000-0001-5450-7067]{Guillaume H\'ebrard}% CONFIRMATION EMAIL
\affiliation{Sorbonne Universit\'e, CNRS, UMR 7095, Institut d’Astrophysique de Paris, 98 bis bd Arago, 75014 Paris, France}

\author[0000-0001-9291-5555]{Charles Cadieux}% CONFIRMATION EMAIL
\affiliation{Universit\'e de Montr\'eal, D\'epartement de Physique, IREX, Montr\'eal, QC H3C 3J7, Canada}

\author[0000-0002-5922-8267]{Thomas Vandal}% CONFIRMATION EMAIL
\affiliation{Universit\'e de Montr\'eal, D\'epartement de Physique, IREX, Montr\'eal, QC H3C 3J7, Canada}

\author[0000-0003-4166-4121]{Neil J. Cook}% CONFIRMATION EMAIL
\affiliation{Universit\'e de Montr\'eal, D\'epartement de Physique, IREX, Montr\'eal, QC H3C 3J7, Canada}

\author[0000-0001-5485-4675]{Ren\'e Doyon}% CONFIRMATION EMAIL
\affiliation{Universit\'e de Montr\'eal, D\'epartement de Physique, IREX, Montr\'eal, QC H3C 3J7, Canada}
\affiliation{Observatoire du Mont-M\'egantic, Universit\'e de Montr\'eal, Montr\'eal H3C 3J7, Canada}

\author[0000-0002-2592-9612]{Jonathan Gagn\'e}% CONFIRMATION EMAIL
\affiliation{Plan\'etarium Rio Tinto Alcan, Espace pour la Vie, 4801 av. Pierre-de Coubertin, Montr\'eal, Qu\'ebec, Canada}
\affiliation{Institute for Research on Exoplanets, Universit\'e de Montr\'eal, D\'epartement de Physique, C.P.~6128 Succ. Centre-ville, Montr\'eal, QC H3C~3J7, Canada}

\author[0000-0002-2842-3924]{Claire Moutou}% CONFIRMATION EMAIL
\affiliation{Univ. de Toulouse, CNRS, IRAP, 14 Avenue Belin, 31400 Toulouse, France}

\author[0000-0002-5084-168X]{Eder Martioli}% CONFIRMATION EMAIL
\affiliation{Sorbonne Universit\'e, CNRS, UMR 7095, Institut d’Astrophysique de Paris, 98 bis bd Arago, 75014 Paris, France}
\affiliation{Laborat\'orio Nacional de Astrof\'isica, Rua Estados Unidos 154, Itajub\'a, MG 37504-364, Brazil}

\author[0000-0002-0474-0896]{Antonio Frasca} % CONFIRMATION EMAIL
\affiliation{INAF - Osservatorio Astrofisico di Catania, Via S. Sofia 78, 95123 Catania, Italy}

\author[0000-0003-0029-2835]{Farbod Jahandar}% CONFIRMATION EMAIL
\affiliation{Universit\'e de Montr\'eal, D\'epartement de Physique, IREX, Montr\'eal, QC H3C 3J7, Canada}

\author[0000-0002-6780-4252]{David Lafreni\`ere}% CONFIRMATION EMAIL
\affiliation{Universit\'e de Montr\'eal, D\'epartement de Physique, IREX, Montr\'eal, QC H3C 3J7, Canada}

\author[0000-0002-8786-8499]{Lison Malo}% CONFIRMATION EMAIL
\affiliation{Universit\'e de Montr\'eal, D\'epartement de Physique, IREX, Montr\'eal, QC H3C 3J7, Canada}
\affiliation{Observatoire du Mont-M\'egantic, Universit\'e de Montr\'eal, Montr\'eal H3C 3J7, Canada}

\author[0000-0001-5541-2887]{Jean-Fran\c cois Donati}% CONFIRMATION EMAIL
\affiliation{Univ. de Toulouse, CNRS, IRAP, 14 Avenue Belin, 31400 Toulouse, France}

\author[0000-0002-6174-4666]{Pia Cortes-Zuleta}% CONFIRMATION EMAIL
\affiliation{Aix Marseille Univ, CNRS, CNES, LAM, Marseille, France}

\author[0000-0001-8388-8399]{Isabelle Boisse}% CONFIRMATION EMAIL
\affiliation{Aix Marseille Univ, CNRS, CNES, LAM, Marseille, France}

\author[0000-0001-5099-7978]{Xavier Delfosse}% CONFIRMATION EMAIL
\affiliation{Univ. Grenoble Alpes, CNRS, IPAG, 38000 Grenoble, France}

\author[0000-0003-2471-1299]{Andres Carmona}% CONFIRMATION EMAIL
\affiliation{Univ. Grenoble Alpes, CNRS, IPAG, 38000 Grenoble, France}

\author[0000-0002-1436-7351]{Pascal Fouqu\'e}% CONFIRMATION EMAIL
\affiliation{Canada-France-Hawaii Telescope, CNRS, Kamuela, HI 96743, USA}
\affiliation{Univ. de Toulouse, CNRS, IRAP, 14 Avenue Belin, 31400 Toulouse, France}

\author[0000-0002-4996-6901]{Julien Morin}% CONFIRMATION EMAIL
\affiliation{Universit\'e de Montpellier, CNRS, LUPM, 34095 Montpellier, France}

\author[0000-0002-5904-1865]{Jason Rowe}% CONFIRMATION EMAIL
\affiliation{Bishops University, 2600 College Street, Sherbrooke, QC J1M 1Z7, Canada}

\author[0000-0001-8134-0389]{Giuseppe Marino}% CONFIRMATION EMAIL
\affiliation{Wild Boar Remote Observatory, San Casciano in val di Pesa, Firenze, 50026 Italy}
\affiliation{INAF - Osservatorio Astrofisico di Catania, Via S. Sofia 78, 95123 Catania, Italy}
\affiliation{Gruppo Astrofili Catanesi - Catania - 95128
Italy}

\author{Riccardo Papini}% CONFIRMATION EMAIL
\affiliation{Wild Boar Remote Observatory, San Casciano in val di Pesa, Firenze, 50026 Italy}
\affiliation{American Association of Variable Star Observers, 49 Bay State Road, Cambridge, MA 02138, USA}
\affiliation{Gruppo Astrofili Catanesi, Catania, 95128, Italy}

\author[0000-0002-5741-3047]{David R. Ciardi} % CONFIRMATION EMAIL
\affiliation{NASA Exoplanet Science Institute-Caltech/IPAC 1200 E. California Blvd, Pasadena CA, 91125, USA}

\author[0000-0003-2527-1598]{Michael B. Lund}% CONFIRMATION EMAIL
\affiliation{NASA Exoplanet Science Institute-Caltech/IPAC 1200 E. California Blvd, Pasadena CA, 91125, USA}

\author[0000-0002-1532-9082]{Jorge H. C. Martins}% CONFIRMATION EMAIL
\affiliation{Instituto de Astrof\'isica e Ci\^encias Espaciais, Universidade do Porto, Rua das Estrelas, Porto, Portugal}

\author[0000-0002-8573-805X]{Stefan Pelletier}% CONFIRMATION EMAIL
\affiliation{Universit\'e de Montr\'eal, D\'epartement de Physique, IREX, Montr\'eal, QC H3C 3J7, Canada}

\author[0000-0002-0111-1234]{Luc Arnold}% CONFIRMATION EMAIL
\affiliation{Canada-France-Hawaii Telescope, CNRS, Kamuela, HI 96743, USA}

\author[0000-0002-7613-393X]{Fran\c cois Bouchy}% CONFIRMATION EMAIL
\affiliation{Departement d’astronomie, Universit\'e de Gen\`eve, Chemin des Maillettes 51, CH-1290 Versoix, Switzerland}

\author[0000-0003-0536-4607]{Thierry Forveille}% CONFIRMATION EMAIL
\affiliation{Univ. Grenoble Alpes, CNRS, IPAG, 38000 Grenoble, France}

\author[0000-0003-4422-2919]{Nuno C. Santos}% CONFIRMATION EMAIL
\affiliation{Instituto de Astrof\'isica e Ci\^encias do Espa\c{c}o, Universidade do Porto, CAUP, Rua das Estrelas, 4150-762 Porto, Portugal}    \affiliation{Departamento de F\'isica e Astronomia, Faculdade de Ci\^encias, Universidade do Porto, Rua do Campo Alegre, 4169-007 Porto, Portugal}

\author[0000-0001-9003-8894]{Xavier Bonfils}% CONFIRMATION EMAIL
\affiliation{Univ. Grenoble Alpes, CNRS, IPAG, 38000 Grenoble, France}

\author[0000-0001-8504-283X]{Pedro Figueira}% CONFIRMATION EMAIL
\affiliation{European Southern Observatory, Alonso de Cordova 3107, Vitacura, Santiago, Chile}
\affiliation{Instituto de Astrof\'{i}sica e Ci\^{e}ncias do Espa\c{c}o, Universidade do Porto, CAUP, Rua das Estrelas, 4150-762 Porto, Portugal}

\author[0000-0002-9113-7162]{Michael~Fausnaugh}
\affiliation{Department of Physics and Kavli Institute for Astrophysics and Space Research, Massachusetts Institute of Technology, Cambridge, MA 02139, USA}

%\author[0000-0003-1447-6344]{Edward~H.~Morgan} 
%\affiliation{Department of Physics and Kavli Institute for Astrophysics and Space Research, Massachusetts Institute of Technology, Cambridge, MA 02139, USA}

%\author{Richard~C.~Kidwell}
%\affiliation{Space Telescope Science Institute, 3700 San Martin Drive, Baltimore, MD, 21218, USA}

\author[0000-0003-2058-6662]{George~Ricker} % CONFIRMATION EMAIL
\affiliation{MIT Kavli Institute for Astrophysics and Space Research, Massachusetts Institute of Technology, Cambridge, MA 02139, USA}
\affiliation{MIT Department of Physics, Massachusetts Institute of Technology, Cambridge, MA 02139, USA}

%\author[0000-0001-6763-6562]{Roland~Vanderspek} # ** no reply **
%\affiliation{MIT Kavli Institute for Astrophysics and Space Research, Massachusetts Institute of Technology, Cambridge, MA 02139, USA}

\author[0000-0001-9911-7388]{David~W.~Latham} % CONFIRMATION EMAIL
\affiliation{Center for Astrophysics $\vert$ Harvard \& Smithsonian, 60 Garden Street, Cambridge, MA 02138, USA}

\author[0000-0002-6892-6948]{Sara~Seager}% CONFIRMATION EMAIL
\affiliation{MIT Kavli Institute for Astrophysics and Space Research, Massachusetts Institute of Technology, Cambridge, MA 02139, USA}
\affiliation{Earth and Planetary Sciences, Massachusetts Institute of Technology, 77 Massachusetts Avenue, Cambridge, MA 02139, USA}
\affiliation{Department of Aeronautics and Astronautics, MIT, 77 Massachusetts Avenue, Cambridge, MA 02139, USA}

\author[0000-0002-4265-047X]{Joshua~N.~Winn}% CONFIRMATION EMAIL
\affiliation{Department of Astrophysical Sciences, Princeton University, 4 Ivy Lane, Princeton, NJ 08544, USA}

\author[0000-0002-4715-9460]{Jon~M.~Jenkins}% CONFIRMATION EMAIL
\affiliation{NASA Ames Research Center, Moffett Field, CA 94035, USA}

\author[0000-0002-8219-9505]{Eric~B.~Ting} % confirmation email
\affiliation{NASA Ames Research Center, Moffett Field, CA 94035, USA}

\author[0000-0002-5286-0251]{Guillermo~Torres}% CONFIRMATION EMAIL
\affiliation{Center for Astrophysics $\vert$ Harvard \& Smithsonian, 60 Garden Street, Cambridge, MA 02138, USA}

%\author{David~Charbonneau}# ** no reply **
%\affiliation{Center for Astrophysics | Harvard & Smithsonian, 60 Garden Street, Cambridge, MA 02138, USA}

%\author[0000-0002-5637-5253]{Alain Lecavelier des \'etangs} # ** declined **
%\affiliation{Institut d’Astrophysique de Paris, UMR7095 CNRS, Universit\'e Pierre & Marie Curie, 98 bis boulevard Arago, 75014 Paris, France}

%\author[0000-0003-3050-8203]{Stanimir Metchev} # ** declined **
%\affiliation{Department of Physics and Astronomy, Centre for Planetary Science and Exploration, The University of Western Ontario, London, ON N6A 3K7, Canada}

\author[0000-0001-8056-9202]{Jo\~ao~Gomes~da~Silva}
\affiliation{Instituto de Astrof\'isica e Ci\^encias do Espa\c{c}o, Universidade do Porto, CAUP, Rua das Estrelas, 4150-762 Porto, Portugal}

\begin{abstract}
We present the discovery of an $18.5\pm0.5$\Mjup brown dwarf (BD) companion to the M0V star \hbox{TOI--1278}. The system was first identified through a percent-deep transit in TESS photometry; further analysis showed it to be a grazing transit of a Jupiter-sized object. Radial velocity (RV) follow-up with the SPIRou near-infrared high-resolution velocimeter and spectropolarimeter in the framework of the 300-night SPIRou Legacy Survey (SLS) carried out at the Canada-France-Hawaii Telescope (CFHT) let to the detection of a Keplerian RV signal with a semi-amplitude of $2306\pm10$\,m/s in phase with the 14.5-day transit period, having a slight but non-zero eccentricity. The intermediate-mass ratio ($M_\star/M_{\rm{comp}} \sim31$) is unique for having such a short separation ($0.095\pm0.001$\,AU) among known M-dwarf systems. Interestingly, M dwarf-brown dwarf systems with similar mass ratios exist with separations of tens to thousands of AUs. 
\end{abstract}

\keywords{near-infrared velocimetry, low-mass stars, brown dwarf}

\section{Introduction}
Over the past 25 years, the search for and characterization of planets around other stars has moved from speculative to one of the most active fields in astronomy. Among planetary systems, those orbiting the coolest stars, M dwarfs, are of particular interest for a number of reasons. They provide  insight into planetary formation in a regime of host star mass that differs significantly from our own Solar System, with some systems more reminiscent in their architecture and scale to the Galilean moon system. Furthermore, as M dwarfs have radii and masses that are significantly smaller than that of our Sun (M0V to M9V dwarfs ranging respectively from 0.6 to 0.1\,R$_\odot$ and 0.6 to 0.1\,M$_\odot$), planets are generally easier  to discover and characterize through transits than similarly sized planets around Sun-like stars. This is particularly true for older, slowly rotating M dwarfs; activity being a limiting factor for a number of low-mass stars. Furthermore, the lower luminosity of M dwarfs (0.1 to $10^{-3.5}$\,L$_\odot$; \citealt{veeder_luminosities_1974,faherty_population_2016}), implies that the habitable zone (HZ) is much closer to the host star than in our Solar System, implying that the orbits of planets in the HZ can be sampled on timescales of days. Given that M dwarfs make up the majority of stars among nearby stellar systems (being five times more numerous than FGK stars) it is likely that most of the HZ planets within 10\,pc are found around M dwarfs \citep{figueira_radial_2016}. This is exemplified by our closest stellar neighbor, the M5.5V Proxima Centauri, having a terrestrial planet in the HZ \citep{anglada-escude_terrestrial_2016}.

As they provide an easier path to the characterization of HZ terrestrial planets, planets around M dwarfs have been the subject of considerable efforts.  Transit searches dedicated to or optimized for M dwarfs \citep{irwin_mearth_2008} have been successful in identifying systems such as the 7 Earth-sized planets around TRAPPIST-1 \citep{gillon_seven_2017}, and the ongoing TESS mission has a top-level requirement of uncovering 1000 small ($R_{\rm p} < 4 $ R$_\oplus$) planets and measuring the masses of 50 such planets \citep{ricker_transiting_2014}.  This requirement is not restricted to M dwarfs, but the TESS bandpass was designed to extend significantly to the red to maximise its sensitivity to planets around M dwarfs, compared to the Kepler mission that had a bluer bandpass better matched to Sun-like stars. While Earth-sized HZ planets receive significant attention, a number of statistical properties of the M dwarf planets have emerged recently regarding more massive companions to M dwarfs.

Jupiter-mass planets are remarkably rare around M dwarfs at short separations. While $0.4-1\%$ of Sun-like stars host a hot Jupiter (occurrence rate estimates differ depending on detection method considered; \citealt{wang_occurrence_2015}), they are rarer around M dwarfs \citep{meyer_m_2017}, although a few systems have recently been discovered (e.g., \citealt{bayliss_ngts-1b_2018}). This is most unlikely to be an observational bias as a hot Jupiter around an M dwarf would induce a $>100$\,m/s radial-velocity signal (from Kepler's laws combining a 1\Mjup on a 4\,days orbit around a 0.1--0.5\Msun host) and a readily detectable transit ($\sim$4\% to $\sim$100\% deep for M0V to M9V; from radius relations in \citealt{boyajian_stellar_2012, filippazzo_fundamental_2015}).

Here, we report the discovery of an $18.5$\,\Mjup transiting brown dwarf (BD) companion on a 14.5--day orbit around the M0V star TOI--1278,  above the maximum mass expected to form through a protoplanetary disk; typically $<1$\,\Mjup with only a few objects between $1$ and $10$\,\Mjup in the 90-star sample by \citet{ansdell_alma_2016}. It occupies a nearly empty part of the mass distribution of close-in companion of M dwarfs, between the Jupiter population and the massive brown-dwarfs companions. In Section~\ref{sec:observations} we present discovery and follow-up observations of the system; in Section~\ref{sec:characterization} we assess the properties of the host star and in Section~\ref{sec:dataanalysis} we detail the data analysis. Finally, in Section~\ref{section:discussion} we discuss the properties of the system in comparison to other planetary and BD companions, its dynamical fate and characterization prospects.

\section{Observations}\label{sec:observations}
We present below the TESS discovery data of TOI--1278, as well as its ground-based follow-up observations in photometry, high-resolution imaging, low-resolution spectroscopy, and precise near-infrared velocimetry, all required to establish the nature of the transiting object TOI--1278\,B.

\subsection{TESS photometry} \label{subsec:tessphotometry}
\begin{figure*}[!htbp]
    \centering
    \includegraphics[width=\linewidth]{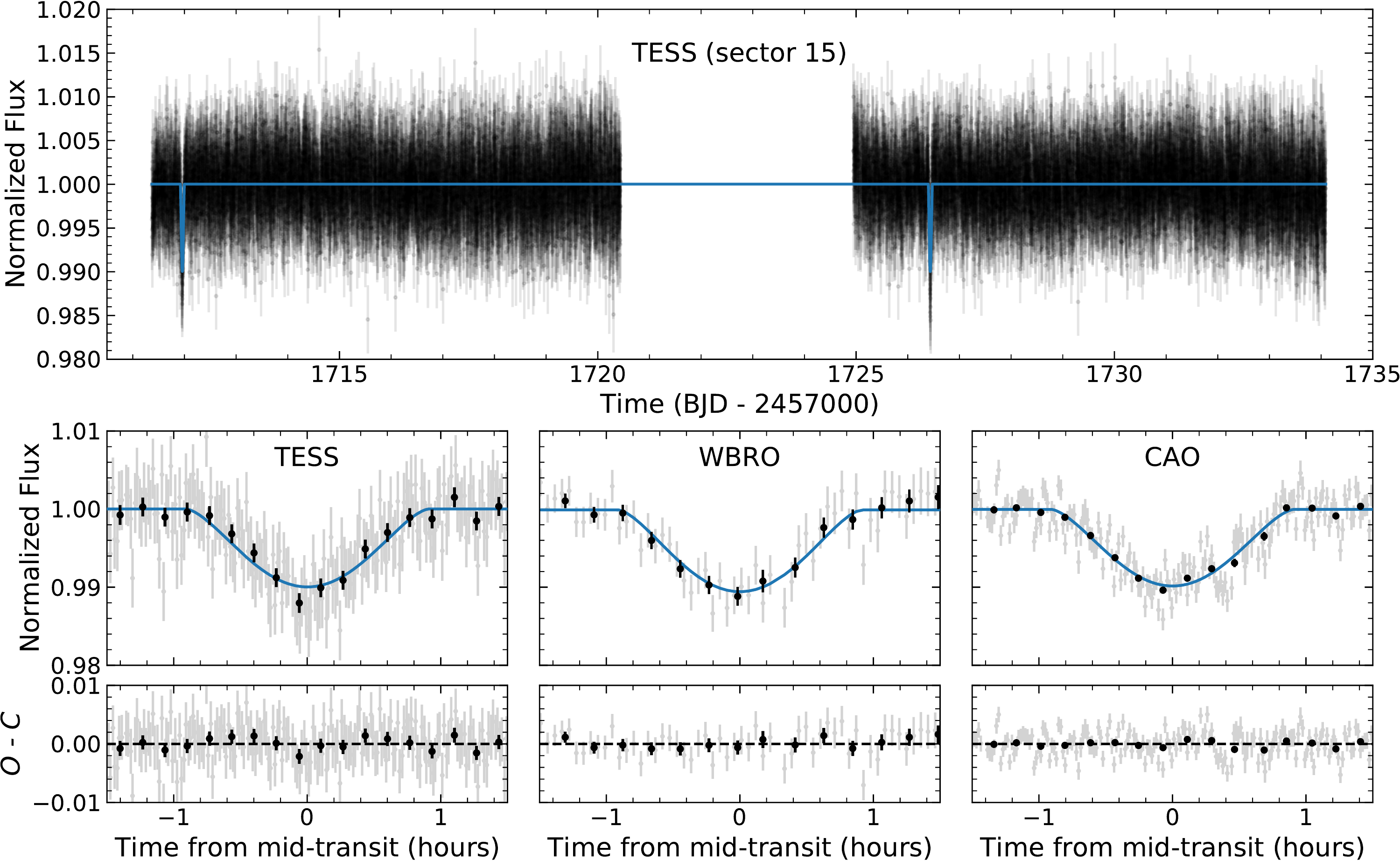}
    \caption{ \textit{Top panel}: TOI--1278 normalized PDCSAP light curve from sector 15. The gap in the middle of the light curve is due to data downlink when TESS is close to perigee. \textit{Lower panels}: TESS, WBRO, and CAO transit light curves. The black points represent binned data (10 min temporal bin). The blue curves are the best-fit transit model, as discussed in Section \ref{sec:dataanalysis}. Only the baseline flux and the limb-darkening coefficients vary between the three data sets. For each instrument, the residuals (Observed - Calculated) are shown below the transit. The TESS data set is phase-folded and covers two transits.}
    \label{fig:lightcurves}
\end{figure*}

TOI--1278 was observed by TESS  \citep{ricker_transiting_2014} in sector 15 with CCD\,4 on Camera\,1 from August 15 to September 11, 2019. This target was selected in the Cool Dwarf List \citep{muirhead_catalog_2018}, a specially curated list of late-K and M dwarfs included in the TESS Input Catalog (TIC; \citealt{stassun_tess_2018}, \citeyear{stassun_revised_2019}) and the Candidate Target List (CTL) for 2-minute cadence light curve sampling. We used the publicly available\footnote{Mikulski Archive for Space Telescopes: \href{https://archive.stsci.edu/tess/}{\nolinkurl{archive.stsci.edu/tess/}}} Presearch Data Conditioning Simple Aperture Photometry (PDCSAP; \citealt{smith_kepler_2012}; \citealt{stumpe_kepler_2012}, \citeyear{stumpe_multiscale_2014}) light curve produced by the NASA Ames Science Processing Operations Center (SPOC; \citealt{chiozzi_tess_2016}). The PDCSAP flux values are corrected for long-term systematic trends seen among other stars in the same sector/camera/CCD using Cotrending Basis Vectors. No long-term stellar variability on timescales shorter than the time series considered here are removed. Furthermore, these PDCSAP fluxes account for the dilution caused by nearby stars. The TIC contamination ratio defined as the flux from other sources within the aperture divided by the target star flux (in TESS-band) is 0.326395 for TOI--1278. 
This dilution correction is important as any other flux contribution would lead to an underestimation of the transiting object radius by measuring a smaller transit depth. TOI-1278 normalized PDCSAP light curve is presented in Figure \ref{fig:lightcurves} and clearly show two transit events.

SPOC sector 15 Data Validation (DV) Reports (\citealt{twicken_kepler_2018}; \citealt{li_kepler_2019}) firstly identified the two transits with a signal-to-noise ratio of 16.5. Both events were reported to be V-shaped, suggestive of a high impact parameter, with an average depth of 1.03$\pm$0.07\,\%. This led to the announcement of planet candidate TOI-1278.01 with an orbital period of 14.476\,days and an estimated radius of 9.3\,R$_\oplus$. Three consecutive transit detections are usually needed to confirm the previous two were produced by the same object. In this case, the two transits were found to have consistent duration, suggesting that they are caused by the same object rather than two planets within the same system or two planets orbiting different stars. We furthermore inspected the PDCSAP light curve near phase $\phi_{\rm sec} = 0.496$, the predicted phase of secondary eclipse based on the geometry of TOI--1278\,B orbit constrained in Section~\ref{sec:dataanalysis}. The eclipse signal due to reflected light is estimated at 13\,ppm for a geometric albedo of 0.5. Given TESS photometric precision, no secondary eclipse is detected and we can only infer a 3-$\sigma$ upper limit of 840\,ppm for the eclipse depth. 

\subsection{Ground-based transit monitoring} \label{subsec:groudbasedphotometry}
While the position of TOI--1278 in the HR diagram (see Section~\ref{sec:characterization}) strongly suggests that the star is not itself an equal-luminosity eclipsing binary, the TESS detection could still be attributed to a nearby eclipsing binary (NEB) located within a few pixels. This is due to TESS's coarse image sampling (21$\arcsec$ per pixel) hence the need to confirm on-target transit events with arc-second angular resolution ground-based observations. In the case of TOI--1278, additional transit observations were crucial to set stronger constraints on its light curve determination, particularly on the orbital period $P$, the radius of the transiting object $R_{\rm p}$ and the impact parameter $b$.  Ground-based monitoring was spurred by the TESS discovery and was obtained to further constrain transit properties. We did not obtain a full out-of-transit light curve.

We scheduled two transit observations using the \texttt{TESS Transit Finder} (\texttt{TTF}), a customized version of the \texttt{Tapir} software package \citep{jensen_eric_tapir_2013}.
We first observed TOI--1278 at Wild Boar Remote Observatory (WBRO) near Florence, Italy on October 27, 2019 without any filter (\textit{clear}). A grazing, percent-deep, transit was detected on-target using a 4$\farcs$7 aperture and was on time given the \texttt{TTF} prediction and uncertainty. The second observation was conducted at {\it M.G. Fracastoro} station (Mt. Etna, 1735 m a.s.l.) of the Catania Astrophysical Observatory (CAO), in Italy on November 25, 2019 in $I_{\rm C}$ with the 0.91-m telescope. Using a 5$\farcs$3 aperture, a second V-shaped, transit of the same depth was detected at the expected time. Both data sets were consistent with TESS in transit depth, shape, duration and timing. Ground-based photometric follow-ups significantly increased the time baseline of our data set and further constrained the orbital fit. We have thus included these additional photometric observations, together with TESS transits, in our joint model described in Section~\ref{sec:dataanalysis}. Calibration and light curve extraction by differential photometry were accomplished using \texttt{AstroImageJ} (\texttt{AIJ}; \citealt{collins_astroimagej_2017}).

\subsection{Ground-based photometry}
To  constrain the stellar parameters of the host star, in particular the effective temperature, we obtained complementary optical photometry at the CAO as the values reported in the TIC had relatively large uncertainties. Photometric measurements in $B$, $V$, $R_{\rm C}$ and $I_{\rm C}$ were obtained on 2020 December 1 (see Table~\ref{tab:properties}). Standard stars in the cluster NGC\,7790 \citep{stetson_homogeneous_2000} were observed just after TOI--1278 to calculate the zero points and the transformation coefficients to the Johnson-Cousins system. The  errors on the $B,V,R_{\rm C},I_{\rm C}$ magnitudes include both the photometric uncertainty coming from the photon statistics and the error of standardization based on the NGC\,7790 data.

\subsection{High-resolution imaging} \label{subsec:highresolutionimaging}
\begin{figure}[!htbp]
    \centering
    \includegraphics[width=\linewidth]{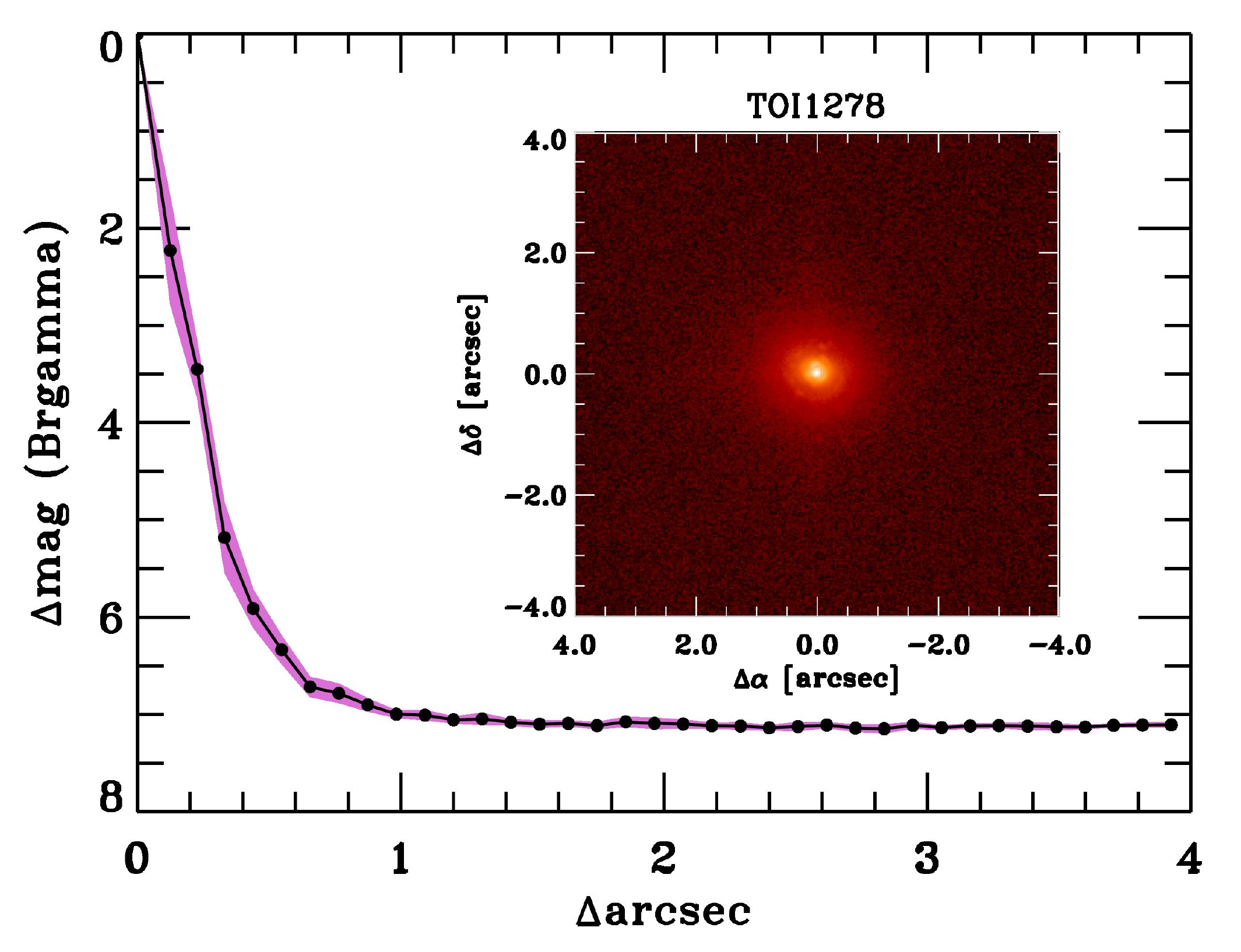}
    \caption{Br$\gamma$ 5$\sigma$ contrast curve of TOI--1278 from Palomar/PHARO NIR Adaptive Optics imaging. No  close companion is detected.}
    \label{fig:Imaging}
\end{figure}

We searched for sources within a few arcseconds of TOI--1278 with Palomar/PHARO NIR Adaptive Optics imaging on November 9, 2019 in the Br$\gamma$ band. As shown in Figure \ref{fig:Imaging}, no companion is detected at 5$\sigma$ with a contrast ratio $\Delta$Br$\gamma \leq$ 7.0\,mag for separation greater than 1$\arcsec$ (76\,AU) and $\Delta$Br$\gamma \leq$ 6.335\,mag for separation over 0.5$\arcsec$ (38\,AU). These contrasts thus exclude a companion brighter than $M_K\sim11.7$, corresponding to an L4 spectral type and $L_{\rm bol}=10^{-4.1}$\,L$_\odot$ for field objects (using polynomial relations in \citealt{faherty_population_2016}). At an age $>1$\,Gyr (see Section~\ref{section:discussion}), this rules-out stellar companion down to the upper limit of the brown dwarf regime (\citealt{phillips_new_2020} models).

\subsection{SPIRou velocimetry} \label{subsec:spirouvelocimetry}

TOI--1278 was observed over 12 epochs from May 31 to November 1, 2020 using SPIRou, the near-infrared ($0.98-2.5\,\mu$m) spectro-polarimeter  installed on the Canada-France-Hawaii telescope \citep{donati_spirou:_2018,donati_spirou_2020}. The data set was obtained as part of the ongoing SPIRou Legacy Survey (SLS; \citealt{donati_spirou_2020}).

At each epoch, a 4 measurement polarimetric sequence was obtained, each measurement having a 5\,min integration, except for June 7 when 8 measurements were obtained due to poorer observing conditions. Per-exposure signal-to-noise ratios (SNR) of {20 to 38} were obtained for the middle of $H$ band. These SNR values are for a spectral sampling of the SPIRou data of $\sim$2.2\,km/s/pixel. We rejected a handful of observations due to low SNR, resulting in 47 usable precision radial velocity (PRV) measurements at 10 different epochs.

Observations were reduced with the standard data reduction pipeline for SPIRou (APERO, version 0.6.131; Cook et al.\,in prep). APERO automatically handles all reduction steps necessary for PRV. In brief, the pipeline first corrects detector-related effects (specific to H4RG infrared arrays, i.e.\ the capacitive coupling between amplifiers, the 1/f noise and the dark current; \citealt{artigau_h4rg_2018}). Then, APERO locates the bad pixel positions, the exact positions of orders on the array and determines the shape of the instrument slicer \citep{micheau_spirou_2018} as well as determining the flat and blaze corrections to apply, using nightly calibration sequences. The calibrated data is then optimally extracted \citep{horne_optimal_1986} in both polarisation channels, simultaneously (AB) and separately (A and B), as well as the simultaneous calibration channel (C). Extracted 2D (E2DS, 49 orders by 4088 pixels) images are produced, and corrected for thermal emission. As the simultaneous Fabry-Perot calibration was used, leakage from the calibration channel into the science channels was also corrected. Nightly wavelength solutions are constructed using a combination of hollow-cathode and Fabry-Perot calibrations (\citealt{cersullo_new_2019}, Hobson et al. in press). Barycentric corrections were determined within APERO using the \texttt{barycorrpy} package \citep{kanodia_python_2018}.

Telluric and night-sky emission correction is also done automatically using APERO. Night-sky emission is corrected using a principal component analysis (PCA) model of OH emission constructed from a library of high-SNR sky observations. Telluric absorption correction is done using a PCA-based approach on residuals after fitting for a basic atmospheric transmission model (TAPAS, \citealt{bertaux_tapas_2014}); the details of the PCA approach are described in \citep{artigau_telluric-line_2014} and its implementation in  APERO  (Artigau et al.\,in prep).

 With the telluric-corrected spectrum, we perform an RV measurement using a cross-correlation function (CCF) with a weighted mask of stellar absorption lines. The mask used must be constructed with a star that reasonably matches in temperature TOI--1278. From existing SPIRou Legacy Survey (SLS) data, we have a number of bright, nearby, M dwarfs that have been observed $\>100$ times with an SNR $>150$ in $H$ band. The best match to TOI--1278 was found to be the M0.5V Gl\,846, and we use the CCF mask constructed from that spectrum through the PRV analysis. 

\begin{figure}
    \centering
    \includegraphics[width=\linewidth]{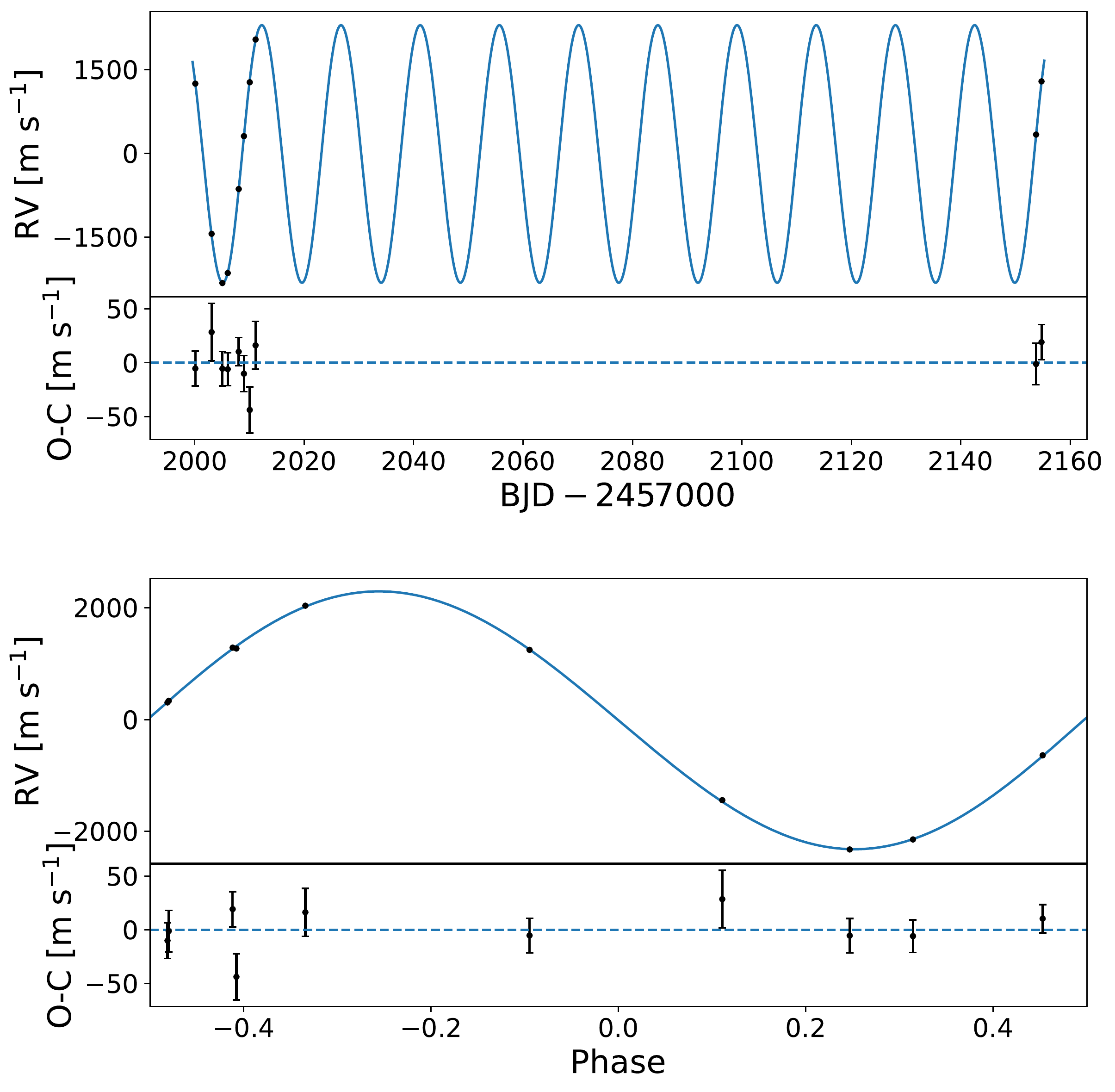}
    \caption{\textit{Top panel:} Radial velocity of TOI--1278 from SPIRou monitoring with an over plot of the best-fit orbital solution. The RV measurements are binned per epoch (typically four observations per night). The median binned error bar is 16.5 \mps{}and the per-epoch residuals (shown below) RMS is 19 \mps{}. The orbital fit is discussed in Section\,\ref{sec:dataanalysis}. \textit{Bottom panel:} Same as the top panel, but phase-folded using the best-fit orbital period and time of inferior conjunction.}
    \label{fig:rv_fit}
\end{figure}

 The CCF is computed for each of the 49 SPIRou orders. Even though the spectra are telluric-corrected, parts of the domain between photometric bandpasses are unusable for PRV. We rejected all orders outside of the $YJHK$ bandpasses. The effective domain used is $980-1113$\,nm, $1153-1354$\,nm, $1462-1808$\,nm and $1957-2400$\,nm. Per-order CCFs for usable orders were weighted by the SNR combined with their radial velocity contents defined in \citealt{bouchy_fundamental_2001} and combined in a per-integration CCF.  We used a cross-correlation of each observation's CCF with the mean CCF of all observations to measure TOI--1278's velocity. This resulted in a slightly lower dispersion of RV than fitting a Gaussian model to each individual CCF. No method that was attempted to measure the velocity from the CCF profile (Gaussian fit, bisector mean position, adjustment of the mean CCF to each observation) led to a significant change in the main result: a $K\sim2300$\,m/s signal consistent in phase and with the period of the companion detected through the photometric monitoring (see Figure \ref{fig:rv_fit}). The per-epoch RV uncertainty was computed by using equations (6) and (13) in \citealt{bouchy_fundamental_2001} for the individual CCFs. The computed uncertainties are consistent with the point-to-point dispersion within each epoch.  To prevent any biases related to under-estimated RV errors, we explored the use of an additional jitter term in the RV model described in Section \ref{sec:dataanalysis}. We found that the posterior of the jitter term was consistent with zero with a $1\sigma$ upper limit at 8 m/s, and that all other model parameters remained unaffected well within their $1\sigma$ uncertainty. We therefore conclude that the reported RV uncertainties are accurate and we do not include the jitter term in the model presented in Section \ref{sec:dataanalysis}.
 
 %{\bf the resulting reduced $\chi^2$ of the orbital fit to the 10 per-epoch measurement show in Figure~\ref{fig:rv_fit} is 1.4, suggesting that the RV uncertainties are only mildly underestimated.}  %Table~\ref{tab:rv} gives per-integration RV values.
 
%\subsection{pRV data}
%\begin{itemize}
%\item Summary of the APERO 0.6.131 used for reduction of pRV data
%\item telluric correction, mostly to say that it's in prep but uses the same ideas as \citep{artigau_telluric-line_2014}.
%\item Standard 'ccf' method, mask constructed from Gl846 data, use only the H+K band as Y+J are really bad in terms of RV content \citep{Artigau2018}.
%\end{itemize}

%\subsubsection{CCF properties} \label{subsec:ccfproperties}
%\begin{itemize}
%\item v sini limits from CCF, comparison with Gl846, the best-fitting M0V in our SPIRou sample
%\item absence of a near-equal luminosity binary and CCF constraints on a blended companion
%\item ccf depth precludes a blended binary
%\end{itemize}

%\begin{figure*}[h]
%    \centering
%    \includegraphics[width=0.95\textwidth]{TOI--1278_mask_gl846_neg_template_CCFs.pdf}
%    \caption{Per-epoch CCFs}
%    \label{fig:RV}
%\end{figure*}

%\begin{figure*}[h]
%    \centering
%    \includegraphics[width=0.95\textwidth]{TOI--1278_mask_gl846_neg_template_residual_CCF.pdf}
%    \caption{Median and residual CCF after correcting for velocity shifts}
%    \label{fig:RV}
%\end{figure*}

\subsection{Near-infrared spectroscopy} \label{subsec:nirspec}

We obtained a low-resolution near-infrared (0.70--2.52\,$\mu$m) spectrum of TOI--1278 on UT 2020 December 24 with the SpeX spectrograph at the NASA Infrared Telescope Facility \citep{2003PASP..115..362R} to improve the wavelength coverage of our spectral energy distribution and refine the estimated radius of TOI--1278. We used the 0\farcs8 slit in prism mode, resulting in a resolving power $\lambda / \Delta\lambda \approx$\,75.

Observing conditions were clear with a 0\farcs5 atmospheric seeing. We obtained four 60--second exposures of TOI--1278 in an ABBA dither pattern along the slit at an airmass of 1.56, followed by standard SpeX prism calibrations. This was immediately followed by the observation of the A0V-type spectral standard HD~203470 to correct for telluric absorption. We obtained thirty short exposures of HD~203470 in ABBA patterns along the slit at an average airmass of 1.64. We used exposure times in the range 0.5--5 seconds to maximize signal-to-noise ratio while ensuring that we would obtain at least one set of exposures inside the linear range of the improved detector response of the recently upgraded SpeX detector (no sequence reached the nonlinear detector regime). 

All data were reduced using \texttt{SpexTool~v4.1} \citep{2004PASP..116..362C}: Two-dimensional traces were first extracted using an optimal extraction algorithm, and each resulting one-dimensional raw spectrum were visually inspected to remove bad pixel clusters that randomly occur in the two-dimensional frames during long exposures. The raw spectra are then combined in a signal-to-noise-weighted average, and the raw science target is corrected for telluric absorption using the standard \texttt{xtellcor} procedure \citep{vacca_method_2003}.

\section{Stellar Characterization}\label{sec:characterization}
\begin{figure}[!htpb]
    \centering
    \includegraphics[width=\linewidth]{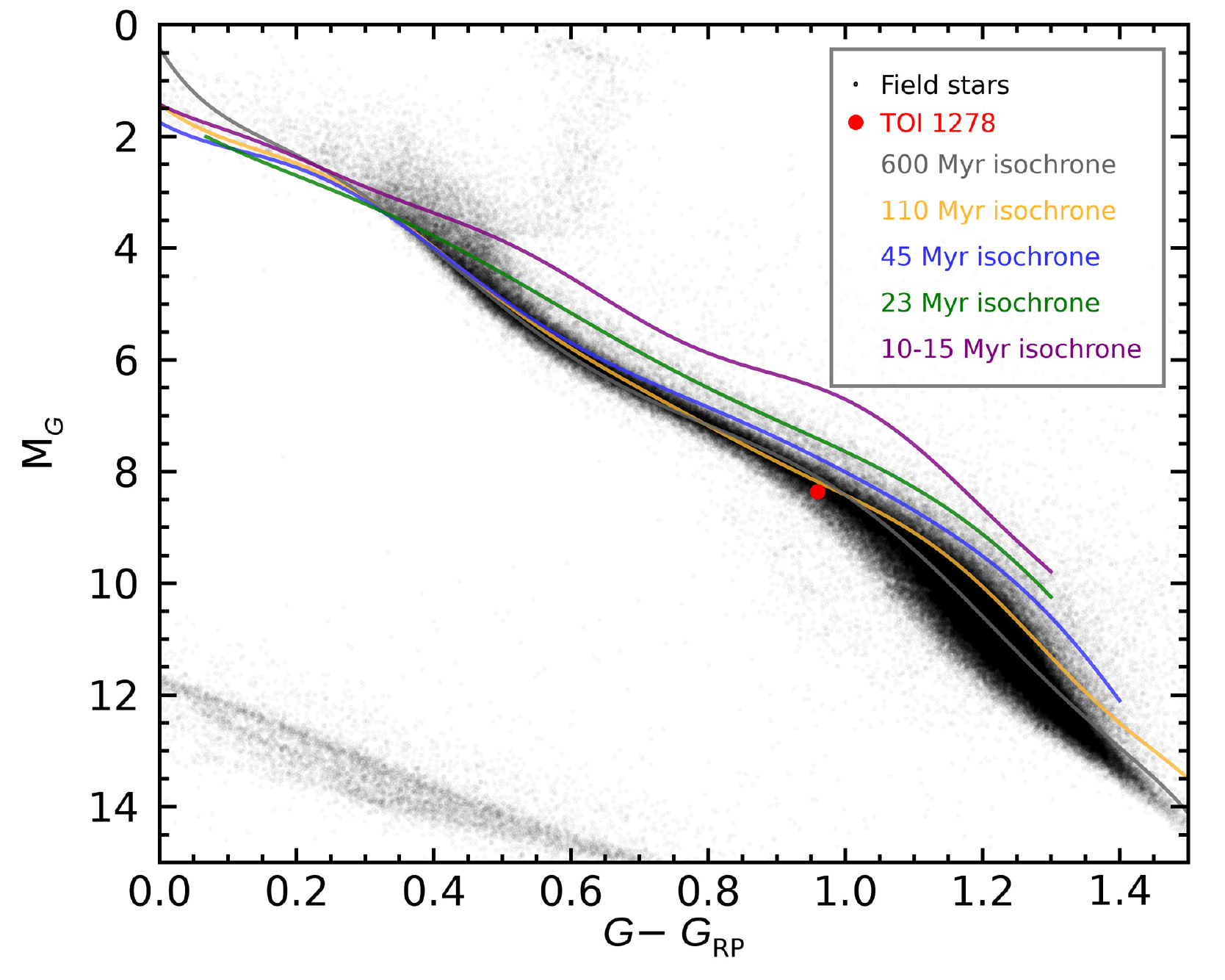}
    %\tablewidth{0.45\textwidth}

    \caption{Position of TOI--1278 for the $<100$\,pc volume-limited sample based on GAIA parallax and photometric measurements. TOI--1278 falls on the main sequence, and it is significantly offset from empirical tracks for young associations that are $<600$\,Myr-old. The position on the HR diagram also rules-out a near-equal luminosity binary, which would be offset from the main sequence by $0.75$\,mag (roughly the position of the 45\,Myr isochrone). Empirical isochrones are from \citet{gagne_banyan_2018} and Gagn\'e et al. (in prep). and are based on the Upper Scorpius, Lower Centaurus Crux and Upper Centaurus associations ($10-15$\,Myr), the $\beta$ Pictoris and 32 Ori associations (23\,Myr), TW Hydra, Columba and Carina associations (45\,Myr), the Pleiades (110\,Myr) and Coma Berenices cluster (600\,Myr).}
    %\label{fig:RV}
    \label{fig:CMD}
\end{figure}

Other than for the presence of a transiting companion (see sections below), \hbox{TOI--1278} is an unremarkable M dwarf at a distance of $75.46\,\pm\,0.07$\,pc \citep{gaia_malin_2020}. In the local-neighborhood HR diagram (see Figure~\ref{fig:CMD}), TOI--1278 falls onto the main sequence of low-mass star, suggestive that it is not an equal-luminosity binary nor particularly inflated due to youth. Also notable is the relatively low level of photometric activity of the star; it displayed a flat light curve, other than for the companion transits, at the part-per-thousand level over the month-long monitoring by TESS (see Figure~\ref{fig:lightcurves}). This argues against a young star that would have frequent flaring; such variability on timescales of hours to days would have been preserved by the PDCSAP photometry.

\subsection{Spectral Energy Distribution and Radius}\label{sec:sed}

\begin{figure}[!htpb]
    \centering
    \includegraphics[width=\linewidth]{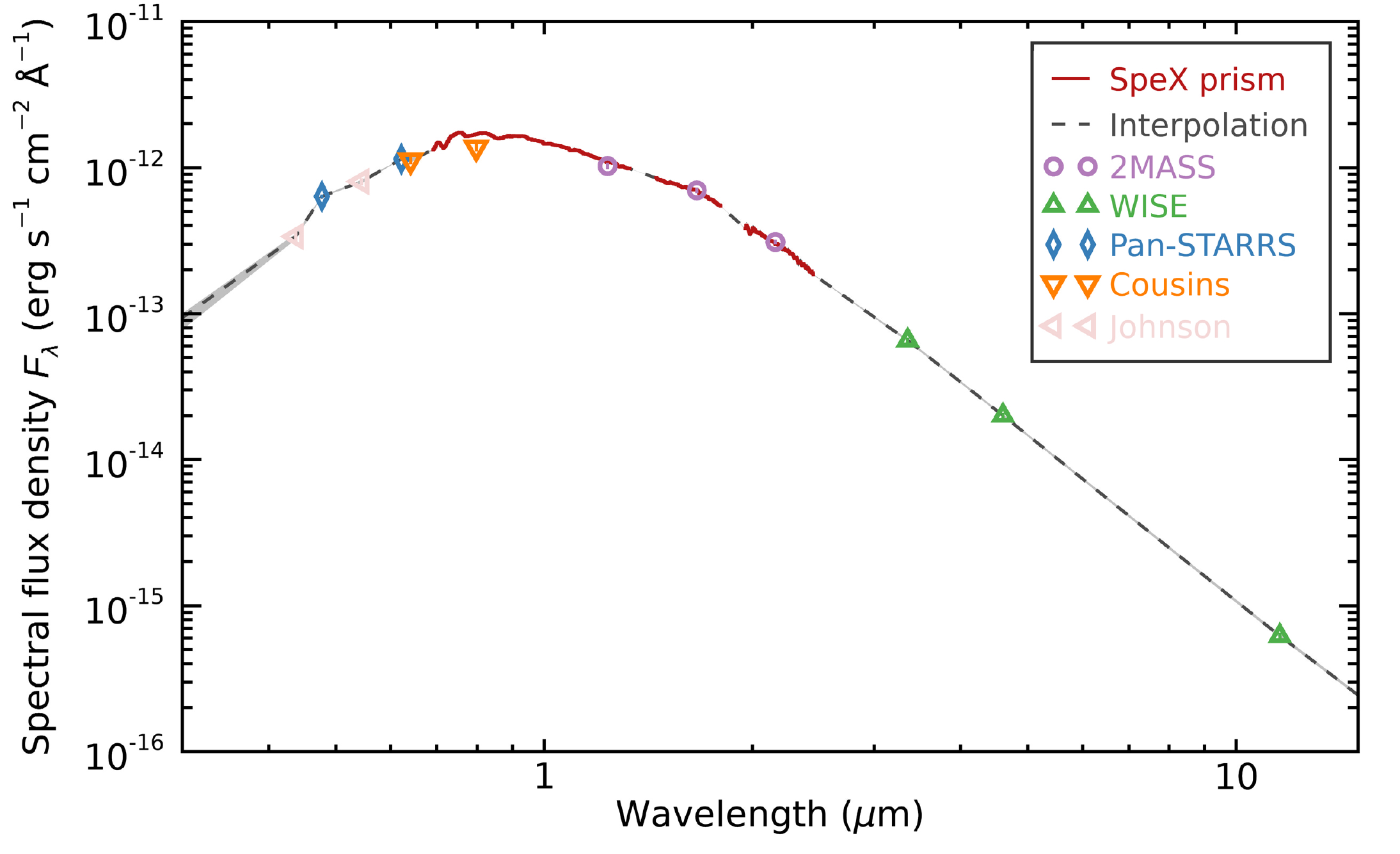}
    %\tablewidth{0.45\textwidth}
    \caption{Spectral energy distribution of TOI--1278. Absolute fluxes estimated using photometry from various surveys are displayed as separate symbols, as indicated in the legend. Our near-infrared SpeX spectrum normalized with absolute fluxes from overlapping photometry is shown as a red solid line. Linear interpolation in logarithm space, and Rayleigh-Jeans and Wien tails of the appropriate effective temperature are shown as dashed lines to complete the spectral energy distribution used in our calculation of the bolometric luminosity.
%The BT-Settl spectrum \citep{allard_model_2010} that provides the best fit to the SED is shown by a grey line.
}
    \label{fig:SED}
\end{figure}

We used the method of \cite{filippazzo_fundamental_2015} to estimate the bolometric luminosity of TOI--1278. We used the updated trigonometric parallax from Gaia~eDR3 to calibrate all available broadband photometry (listed in Table~\ref{tab:properties}), and our SpeX near-infrared spectrum was normalized to an error-weighted combination of the predicted absolute fluxes from the broadband photometric measurements that overlap with the spectrum, following \cite{filippazzo_fundamental_2015}. We adopted the effective temperature determined in Section~\ref{subsection:stellarparameters} ($3799 \pm 42$\,K) and assigned a Rayleigh-Jeans and Wien tails of the appropriate blackbody temperature. Linear interpolation in logarithm space was used to construct the spectral energy distribution outside of the SpeX spectrum and between available broadband photometric measurements. Our resulting spectral energy distribution, shown in Figure~\ref{fig:SED}, allowed us to measure a bolometric luminosity of $0.0614 \pm 0.0001$\,$\textrm{L}_\odot$.  This bolometric luminosity measurement is semi-empirical; one only needs to assume a temperature to set the shape of the Wien tail. Considering that it accounts for a small fraction of the overall flux, the impact of the uncertainty on the effective temperature is much smaller than the uncertainties propagated from photometry and parallax measurements. Regarding the Rayleigh-Jeans regime, the only assumptions is that the two reddest photometric data points are well past the peak of the the Plank function, free of strong molecular absorption and not significantly affected by a circumstellar disk.  Considering that the SED peaks at $\sim1$\,$\mu$m and that the reddest bandpasses are at 4.6 and 11.6\,$\mu$m, this assumption is justified. Combining this measurement with our model-dependent $T_{\rm eff}$ allows us to derive a semi-empirical radius measurement of $0.573\pm0.012$\,\Rsun. This measurement is marginally larger (by $5 \pm 4$\%) than predictions from the solar-metallicity models of \cite{baraffe_evolutionary_1998}, and only $1 \pm 3$\% larger than empirical spectral type to radius relations for field stars \citep{pecaut_intrinsic_2013}\footnote{See also \href{http://www.pas.rochester.edu/~emamajek/EEM_dwarf_UBVIJHK_colors_Teff.txt}{\nolinkurl{pas.rochester.edu/~emamajek/EEM_dwarf_UBVIJHK_colors_Teff.txt}}}. This radius measurement is also consistent with interferometric measurements for nearby M dwarfs \citep{demory_mass-radius_2009}, which tend to be marginally inflated compared with solar-metallicity models, by about 0--2\%. This is in stark contrast with young M dwarfs, which tend to be significantly inflated (by 150--250\% at 20--40\,Myr) compared with theoretical models and older M dwarfs, likely because of their faster rotation rates that drive strong magnetic fields which in turn affect the hydrostatic balance of young M dwarfs \citep{malo_banyan._2014}. This indicates that TOI--1278 is likely not a young M dwarf, consistent with our small $v\sin i$ measurement (see Section \ref{subsection:stellarparameters}).

Our SpeX low-resolution near-infrared spectrum indicates an M0 spectral type when compared with spectral type standards in the IRTF spectral library \citep{cushing_infrared_2005,rayner_infrared_2009}, consistent with our best-matching CCF mask from the M0.5V-type template Gl846. Constraining the age of the TOI--1278 is key for further analysis of its properties; a young M0V  would have an inflated radius and lower mass compared to Gyr-old objects. This would impact both the derived mass from radial velocity measurements and radius estimate from transit properties. Also, uncovering a young transiting brown dwarf would be  interesting to constrain evolution models.

Rare objects such as the young ($93^{+61}_{-29}$\,Myr) inflated transiting brown dwarf TOI-811\,b provide complementary constraints to brown dwarf evolutionary models \citep{carmichael_toi-811b_2020}. This provides a complementary constraint compared to objects of similar mass and ages in young moving groups for which the radius is constrained through bolometric properties \citep{filippazzo_fundamental_2015}.

Young brown dwarfs have a significant self-luminosity, and a  young system would be significantly easier to detect as a double-lined spectroscopic binary. Furthermore, the TOI--1278 system could consist of an unresolved binary system with the transiting companion around one of the two components. While there are a number of well established tools to rule out a number of scenarios for transiting companions (see Sections~\ref{subsec:tessphotometry} and \ref{subsec:highresolutionimaging}), simple color-magnitude diagram analysis can already exclude a number of configurations.

\begin{deluxetable}{lc}
\tablenum{1}
\tablecaption{Basic data of TOI--1278\label{tab:parallax}}
\tablehead{
\colhead{Quantity} & \colhead{Value} 
}
\startdata
\multicolumn{2}{c}{Designations}\\
\multicolumn{2}{l}{WISEA\,J212154.86+353855.3}\\
\multicolumn{2}{l}{Gaia EDR3 1867491607542113536}\\
\multicolumn{2}{l}{2MASS\,J21215495+3538557}\\
\multicolumn{2}{l}{TIC\,163539739}\\ 
\multicolumn{2}{l}{TOI--1278}\\ \hline
\multicolumn{2}{c}{Astrometry}\\
RA & $320.47842885720$\tna \\
DEC & $+35.64861839399$\tna\\
Parallax & $13.25\pm0.01$\tna \\
$\mu_\alpha$ [mas/yr] & $-89.474\pm0.010$\tna\\
$\mu_\delta$ [mas/yr] & $-46.944\pm0.011$\tna\\ \hline
\multicolumn{2}{c}{Galactic kinematics}\\
X & $10.208\pm0.009$\,pc\tnb \\
Y & $73.58\pm0.07$\,pc\tnb \\
Z & $-13.26\pm 0.01$\,pc\tnb \\
U & $30.29\pm0.04$\,km/s\tnb\\
V & $-31.9\pm0.2$\,km/s\tnb\\
W & $16.28\pm0.04$\,km/s\tnb\\ \hline
\enddata
\tablecomments{\tna Gaia EDR3 \citep{gaia_malin_2020}, \tnb this work, combining GAIA astrometry and SPIRou velocimetry.}
\end{deluxetable}

\begin{deluxetable}{lc}
\tablenum{2}
\tablecaption{Bulk properties of TOI--1278\label{tab:properties}}
\tablehead{
\colhead{Quantity} & \colhead{Value} 
}
\startdata
\multicolumn{2}{c}{ Stellar parameters}\\
Mass (\Msun) & $0.55\pm0.02$\tna\\           % \citealt{stassun_tess_2018}
Radius (\Rsun) & $0.573\pm0.012$\tnb\\        % this work
 log $g$ & $4.68\pm0.10$\tnb\\                     % this work
Luminosity (L$_\odot$) & $0.0614\pm0.0001$\tnb\\            % this work
$T_{\rm eff}$ (K) & $3799\pm42$\tnb\\            % this work
$v$ sin $i$ (km/s) & $1.10\pm0.86$\tnb\\         % this work
A$_V$ (mag) & $-0.01\pm0.10$\tnb \\     % this work
 \feh& $-0.01\pm0.28$\tnb \\    \hline  
Spectral type & M0V\tnb \\  \hline                % \citealt{stassun_tess_2018}
\multicolumn{2}{c}{ Photometry }\\
$B$ & $15.07\pm0.05$\tnb\\      % this work
$V$ & $13.51\pm0.03$\tnb\\      % this work
$R_c$ & $12.60\pm0.04$\tnb\\      % this work
$I_c$ & $11.70\pm0.04$\tnb\\      % this work
$G$ & $12.740\pm0.003$\tnc \\      %\citep{gaia_malin_2020},
$g$ & $14.078\pm0.007$\tnd\\      %Pan-STARRS1 \citep{chambers_pan-starrs1_2019}
$r$ & $12.898\pm0.001$\tnd\\      %Pan-STARRS1 \citep{chambers_pan-starrs1_2019}
$i$ & $12.30\pm0.04$\tnd\\      %Pan-STARRS1 \citep{chambers_pan-starrs1_2019}
$z$ & $12.5\pm0.4$\tnd\\      %Pan-STARRS1 \citep{chambers_pan-starrs1_2019}
$y$ & $12.07\pm0.05$\tnd\\      %Pan-STARRS1 \citep{chambers_pan-starrs1_2019}
$J_{2M}$ & $10.601\pm0.025$\tne\\        %2MASS \citep{skrutskie_two_2006}
$H_{2M}$ & $9.939\pm0.021$\tne\\        %2MASS \citep{skrutskie_two_2006}
$K_{\rm s,2M}$ & $9.735\pm0.012$\tne\\        %2MASS \citep{skrutskie_two_2006}
$W1$ & $9.629\pm0.022$\tnf\\              %All-WISE \citep{cutri_r_m_et_al_vizier_2014}
$W2$ & $9.591\pm0.020$\tnf\\              %All-WISE \citep{cutri_r_m_et_al_vizier_2014}
$W3$ & $9.443\pm0.03$\tnf\\              %All-WISE \citep{cutri_r_m_et_al_vizier_2014}
\hline
\enddata
\tablecomments{
\tna \citealt{stassun_tess_2018},
\tnb this work,
\tnc Gaia EDR3 \citep{gaia_malin_2020}, 
\tnd \hbox{Pan-STARRS1} \citep{chambers_pan-starrs1_2019}, 
\tne 2MASS \citep{skrutskie_two_2006}, 
\tnf AllWISE \citep{cutri_r_m_et_al_vizier_2014}.
}
\end{deluxetable}

\subsection{Stellar parameters from SPIRou spectra}\label{subsection:stellarparameters}

The determination of the atmospheric parameters (\teff\ and \logg) and projected rotational velocity (\vsini) was accomplished through the code \texttt{ROTFIT} \citep[e.g.,][]{frasca_x-shooter_2017}. 

We adopted a grid of synthetic BT-Settl spectra as templates \citep{allard_model_2010} at $[Fe/H] = -0.5, 0.0$ and $0.3$, and effective temperature in the range \hbox{3000--4500\,K} (in steps of 100\,K) and \logg\ from 5.5 to 2.5 dex (in steps of 0.5 dex).

\begin{figure}
\centering
\includegraphics[width=4.7cm]{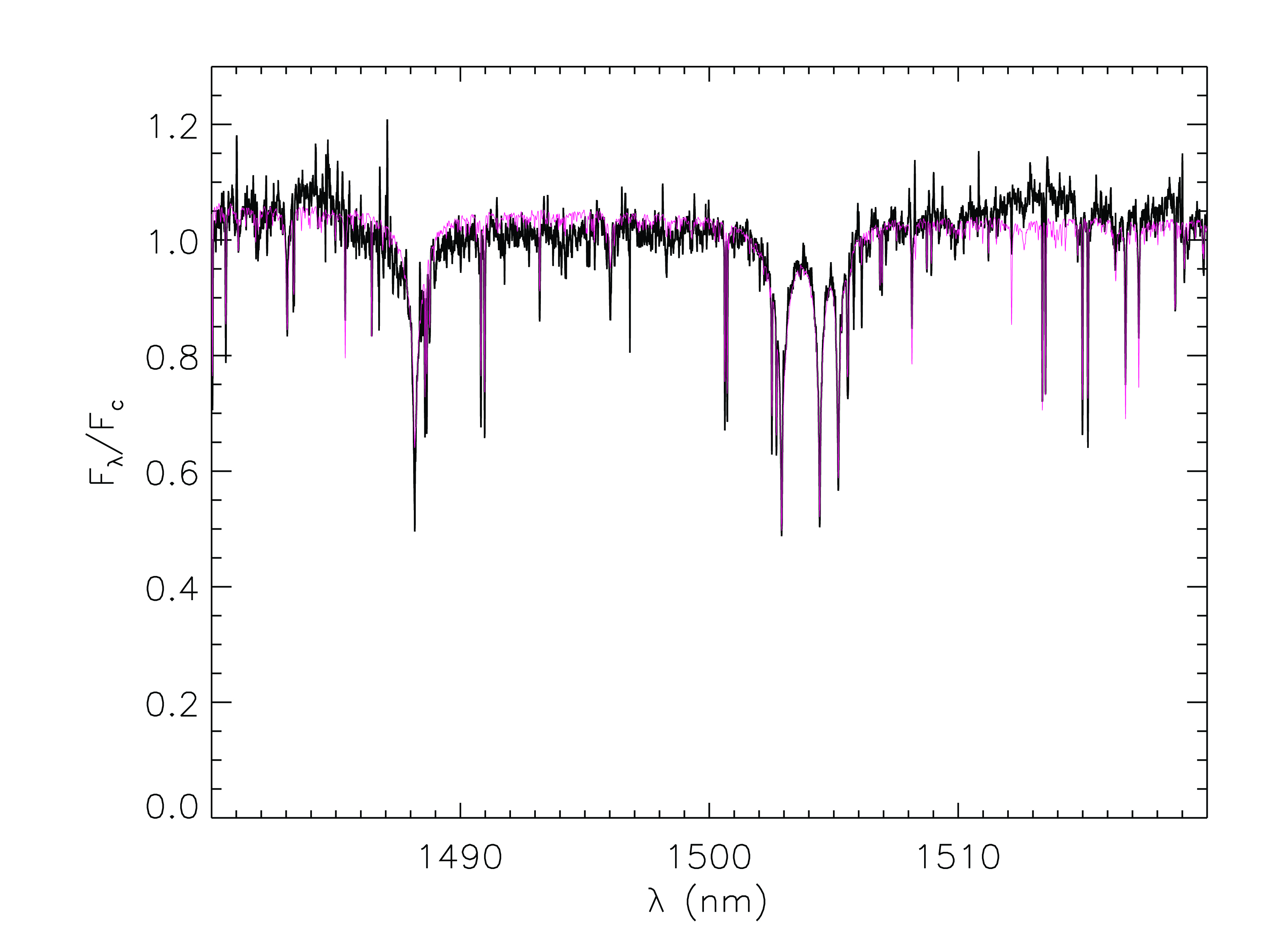}
\includegraphics[width=3.7cm]{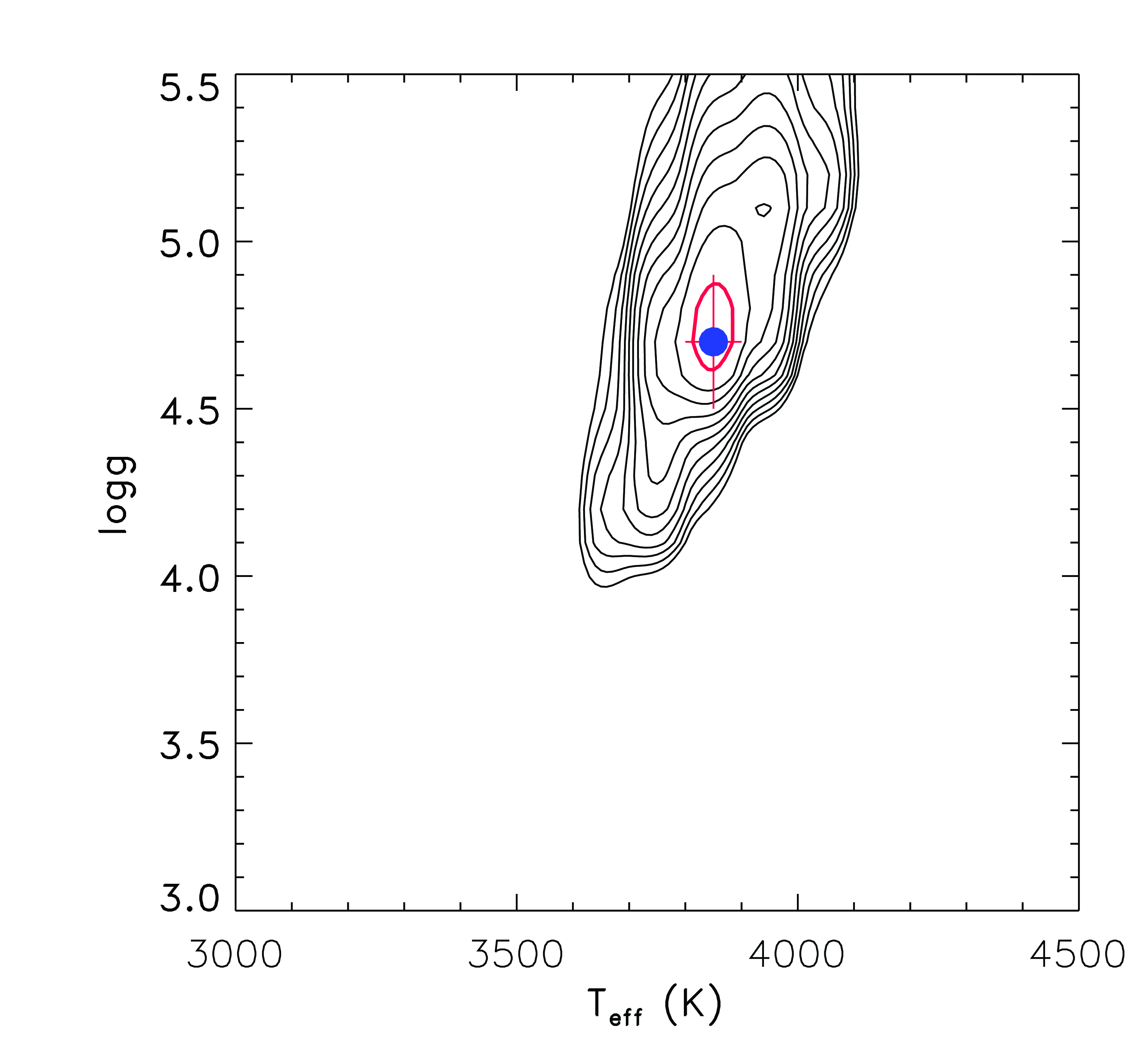}
\includegraphics[width=4.7cm]{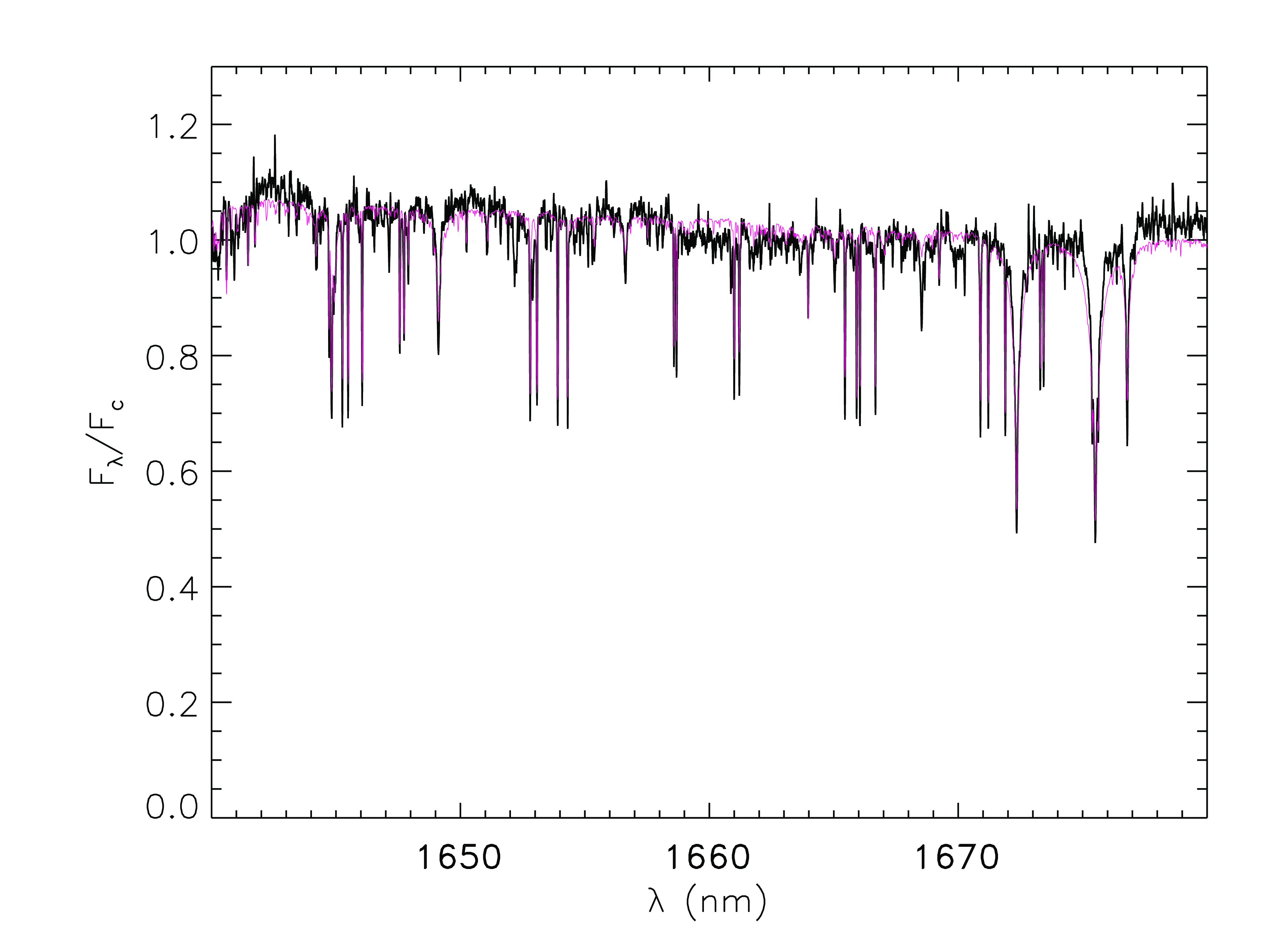}
\includegraphics[width=3.7cm]{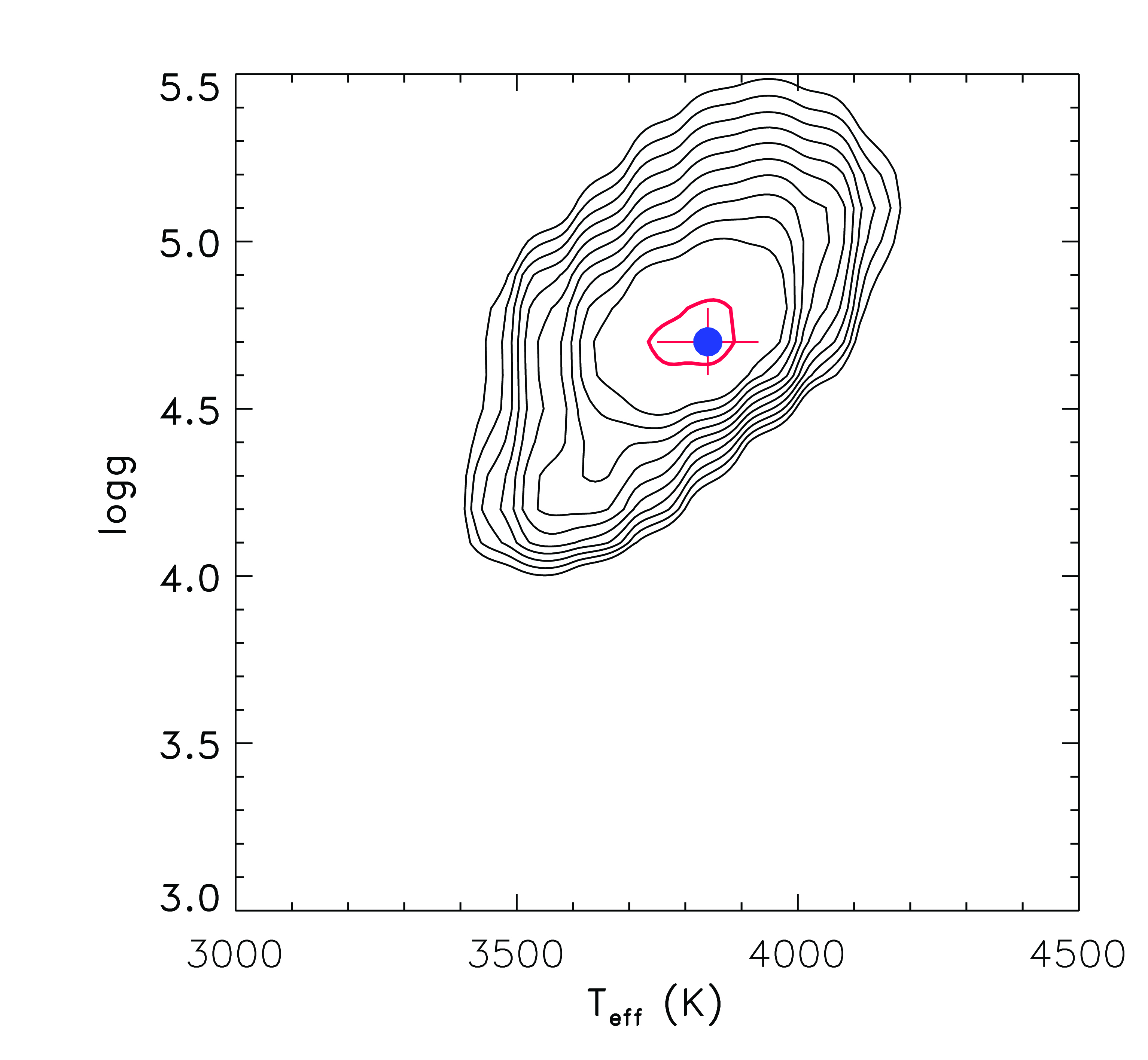}
\includegraphics[width=4.7cm]{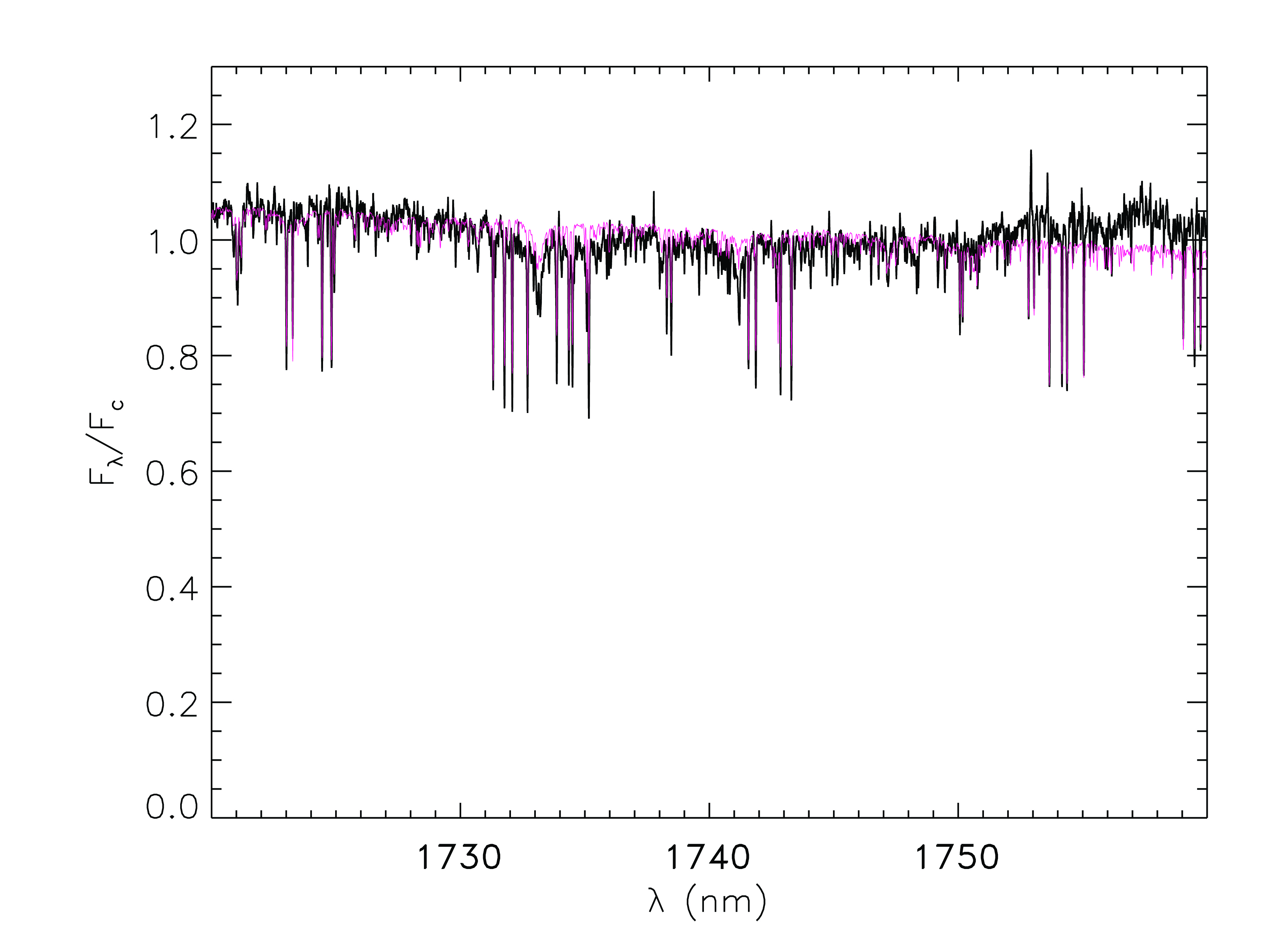}
\includegraphics[width=3.7cm]{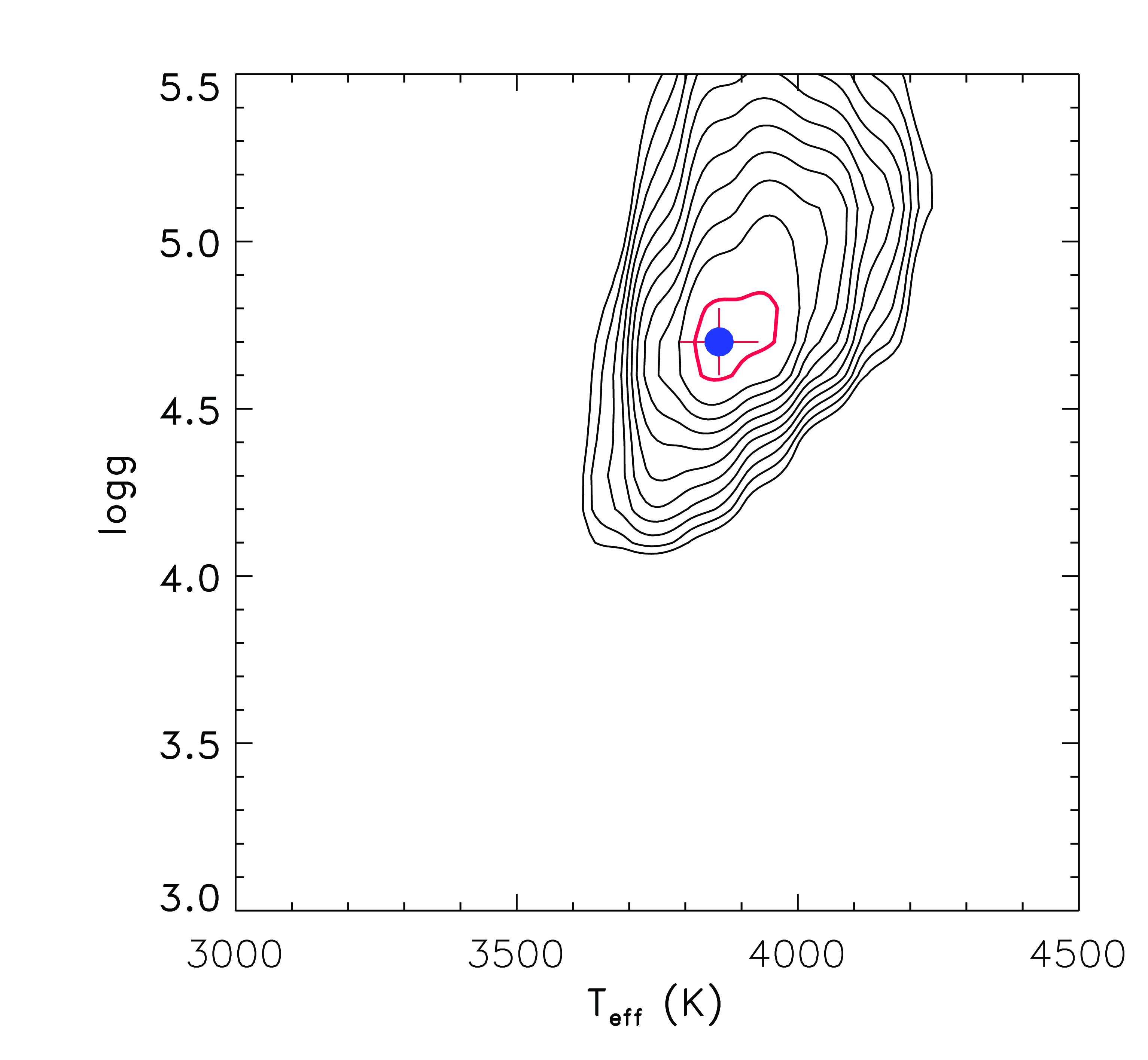}
\vspace{0cm}
\caption{{\it Left panels:} continuum-normalized co-added SPIRou spectrum of TOI--1278 in three regions (black full lines) with the best fitting synthetic spectrum  
over-plotted (red dotted lines). {\it Right panels:} $\chi^2$ contour maps in the \teff-\logg\ plane. In each panel, the 1$\sigma$ confidence level is 
denoted by the red contour. The best values and error bars on \teff\ and \logg\ are also indicated. }
\label{fig:ROTFIT}
\end{figure}

\texttt{ROTFIT} finds the best values of atmospheric parameters and \vsini\ by minimizing 
the $\chi^2$ of the difference between the observed and synthetic spectra in specific spectral segments.  We used the $H$ band (eight 40\,nm-wide segments) as well as regions centered on strong lines within $J$ and $K$ (1290-1330\,nm and 1960-2000\,nm)  for the  \texttt{ROTFIT} analysis. These domains avoid the stronger telluric absorption bands between photometric bandpasses that may display significant correction residuals, avoids the increased thermal noise on the red side of $K$ band. Some regions within $Y$ and $J$ are relatively clean of tellurics, but have fewer strong and sharp spectroscopic features than the regions selected for the analysis; this manifest itself in the very poor radial-velocity content of $Y$ and $J$ band for M dwarfs (e.g., \citealt{figueira_radial_2016, artigau_optical_2018}). Figure~\ref{fig:ROTFIT}  shows sample regions for the fit on the telluric-corrected co-added SPIRou spectrum.  Both templates and observed spectra are normalized to the local continuum and the templates are degraded to the SPIRou spectral resolution and resampled on the points of the target spectrum. The templates are also artificially broadened by  convolution with a rotational profile whose  \vsini\  span a range of values to find the minimum of $\chi^2$. For each spectral segment, the best values of \teff\ and \logg\ are found by interpolation in the grid of templates, as shown in the right panels of Fig.~\ref{fig:ROTFIT}.

% Updated values ::
%# <vsini>      1.10 +/-   0.69 +/-   0.86
%# <Teff>    3799.03 +/-  19.40 +/-  42.16
%# <logg>       4.68 +/-   0.04 +/-   0.10

 The values of  \teff\, = $3799\pm42$\,K, \logg\, = $4.68\pm0.10$\,dex, and \vsini\, = $1.10\pm 0.86$\,km/s are the weighted means of the values from each $i$-th spectral region, where the weights are $w_i=1/\sigma_i^2$ ($\sigma_i$ is the error of the parameter for the $i$-th segment. The analysis was performed at 3 metallicity values, and the lowest $\chi^2$ value was obtained for $[Fe/H] = 0.0$. The uncertainty of each parameter is the largest between the standard error of the weighted mean and the weighted standard deviation. The value of \vsini \ is significantly smaller than the spectral resolution of SPIRou ($\sim4.3$\,km/s), and is sensitive to other broadening effects, such as the macro-turbulence, that are not  distinguishable from rotation at low \vsini\ values. The \vsini\ should therefore be taken as a very approximate value.

% This is not exactly what happens in the BT-Settl spectra. In the NIR region, from 1300 to 2000 nm the sampling is practically constant ~0.20 A/pixel, EXCEPT within the strongest lines where it can be as low as 0.04-0.05 A.
%I would speak about sampling rather than resolution... e.g.
% The sampling of BT-Settl models is not uniform in the SPIROU spectral range. It is typically 0.20 A, but it can be as low as 0.05, which is comparable with that of SPIROU, inside strong spectral lines. 

\subsection{ Refining metallicity measurements\label{feh_determination}}

The high-resolution spectra of TOI--1278 let us determine the metallicity of the star with comprehensive examinations of individual  \hbox{Fe I} lines in all $Y$, $J_{2M}$, $H_{2M}$ and $K_{2M}$ bands.  BT-Settl and ACES models \citep{allard_stellar_2011, husser_new_2013} are available at varying resolution over the SPIRou domain. The sampling of BT-Settl models is not uniform; it is typically 0.20\,\AA, but it can be as low as 0.05\,\AA, which is comparable with that of SPIRou, inside strong spectral lines. The determination of \teff \ and {\logg} discussed in the section above is not strongly sensitive to the  resolution, but in contrast, metallicity measurements are more strongly sensitive to the data resolution. We therefore  decided to use PHOENIX ACES models, available at a resolution higher than that of SPIRou, for metallicity measurements of all identified individual Fe I absorption lines.
To do this we generated multiple synthetic spectra for the fixed effective temperature and {\logg} of 3800\,K and 4.5 dex respectively and the metallicity range of $-1.5$ to $0.5$\,dex. The \teff\,  and {\logg} values were adopted from \texttt{ROTFIT} results (see Table \ref{tab:properties}) and then rounded to the nearest 100\,K and 0.5\,dex.

%{\textbf{Note that we have empirically confirmed that offsets up to 150\,K for \teff and 0.5 for {\logg} from the predetermined values have a negligible effect on abundance measurements, and practically the uncertainty caused by these offsets is dominated by the larger uncertainty of line measurements}

To match the original resolution and sampling of the ACES models to that of the observed data, we degraded and convolved all the generated synthetic data with respect to a SPIRou spectrum. Then we used the Vienna Atomic Line Database (VALD; \citealt{ryabchikova_major_2015}; \citealt{kupka_vald-2_2000}; \citealt{piskunov_vald_1995}) to identify all visible Fe I lines on the models. Finally, we performed $\chi^2$ analyses between the data and the models on each line individually. Note that in total, 78 Fe\,I lines were detected and used in our analysis. The result of this chemical spectroscopy is metallicity of \feh\,$= -0.01\pm0.28$, where the conservative uncertainty is directly from the standard deviation of all the measurements. Hence we conclude that TOI--1278 has a Solar metallicity, in agreement with the ROTFIT results.

\subsection{Spectropolarimetry with SPIRou}
\begin{figure}
\centering
\includegraphics[width=0.99\linewidth]{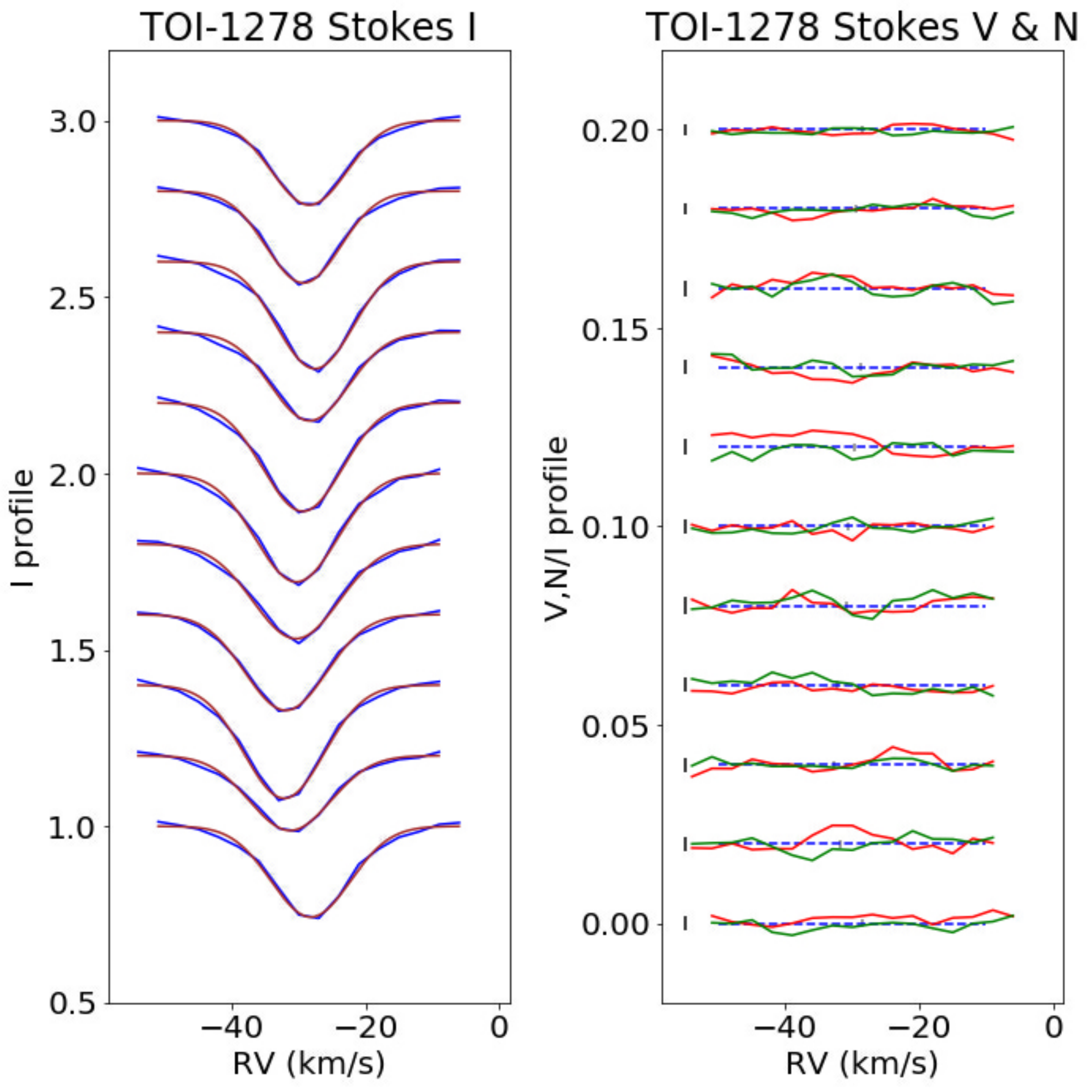}
\vspace{0cm}
\caption{Stokes I (left) and Stokes V (right) of all polarization sequences obtained on TOI–1278. The error bars close to Stokes V profiles show the mean uncertainty on each data point. The green curves on the right plot show the Null check profiles. Superimposed to the intensity profiles is a gaussian fit.}
\label{fig:POLAR}
\end{figure}

Given that each visit to TOI--1278 is composed of a polarimetric sequence, it is possible to combine the individual spectra of each beam to have access to the circular polarization of the stellar surface \citep{donati_spectropolarimetric_1997, donati_spirou_2020}. Using Least-Square Deconvolution and a line list, a mean profile is obtained for the intensity spectrum (Stokes I profile) as well as for the polarized spectrum (Stokes V profile). The line list used for TOI--1278 is based on VALD database for an effective temperature of 3750 K and minimum line depth of 0.15 of the continuum; it contains 500 atomic lines. The description of the data analysis method as included in the APERO pipeline is given in \citet{martioli_spin-orbit_2020}. Figure \ref{fig:POLAR} shows the resulting Stokes I and Stokes V profiles for the whole data set. The noise level in the Stokes V profile corresponds to a few 10$^{-4}$ of the intensity spectrum, with no significant detection in the available time series. Using the relationship between the mean Land\'e factor, the mean wavelength and the Stokes V and I profiles as introduced in \cite{donati_spectropolarimetric_1997}, one can find a 1-$\sigma$ upper limit of 22 to 35 Gauss for the longitudinal field on these visits.

\section{Data analysis \& Results} \label{sec:dataanalysis}
%In Section \ref{subsection:jointanalysis}, we present a joint analysis of the three light curves and the SPIRou RV by modelling all data simultaneously.
%\subsection{Joint pRV and Photometric Analysis}\label{subsection:jointanalysis}

To constrain the orbital and physical parameters of TOI--1278\,B, we performed a joint analysis of the SPIRou RV data along with the TESS and ground-based transits. We used the Python package \texttt{exoplanet} \citep{foreman-mackey_exoplanet-devexoplanet_2020}, which computes analytical transit models using \texttt{starry} \citep{luger_mathttsmathtttmathttmathttrmathttrmathtty_2019} and solves the Kepler equation with a built-in solver. The \texttt{exoplanet} framework uses \texttt{PyMC3} \citep{salvatier_probabilistic_2015} to generate Bayesian models and perform gradient-based Markov Chain Monte Carlo (MCMC). Models in \texttt{PyMC3} are built with the \texttt{Theano} \citep{al-rfou_theano_2016} numerical infrastructure.

With this framework, we modeled the four transits (two from TESS, one from WBRO and one from CAO) and the full SPIRou RV time series simultaneously. We included a separate baseline flux $f_0$ and quadratic limb-darkening coefficients $\left\{u_1, u_2\right\}$ for each instrument. For TESS, the coefficients are determined by interpolating Table 5 of \citet{claret_new_2018} given $T_{\rm eff}$ and $\log g$ from Table \ref{tab:properties}, and are held fixed to $\left\{0.2336, 0.3710\right\}$. For CAO, we used EXOFAST \citep{eastman_exofast_2013} to interpolate tables from \citet{claret_gravity_2011}\footnote{See also \href{http://astroutils.astronomy.ohio-state.edu/exofast/limbdark.shtml}{\nolinkurl{astroutils.astronomy.ohio-state.edu/exofast/limbdark.shtml}}}. We assumed $[\textrm{Fe/H}] = 0$ (see Section \ref{feh_determination}) and fixed the resulting $I_{\rm c}$ band-pass coefficients to $\left\{0.2776, 0.3528\right\}$. Since the WBRO transit was observed without any filter, we chose to let $\left\{u_1, u_2\right\}$ vary following a Gaussian distribution $\mathcal{N}(0.3, 0.1)$. 

We used relatively narrow uniform priors on the period $P$ and the time of mid-transit $T_0$, but their range are still several orders of magnitude larger than the uncertainties from the preliminary analysis by the TESS team (DV: $P = 14.4762 \pm 0.0022$ days, $T_0 = 2458711.959717 \pm 0.0015$ BJD). We also imposed Gaussian priors on the stellar radius $R_{\rm s}$ and mass $M_{\rm s}$ using values from Table \ref{tab:properties}. We used uniform priors on the orbital eccentricity $e$ and argument of periastron $\omega$, as well as for the systemic velocity $\gamma$. For the RV semi-amplitude $K$, we used a broad log-uniform prior to sample several orders of magnitude uniformly. The transit impact parameter $b$ has a uniform prior between 0 and $1+R_{\rm p}/R_{\rm s}$. Since the transit is grazing, a strong degeneracy exists between $b$ and the planetary radius $R_{\rm p}$. To alleviate this degeneracy and assess its importance, we compared three different priors on $R_{\rm p}$: a broad uniform prior between 0.0 and 2.5\,\Rjup{}, a Gaussian prior based on the latest evolutionary models described in Section~\ref{subsection:apeculiartemperatureregime}, and a Gaussian prior derived from the confirmed exoplanet population. To construct the latter, we follow an approach similar to \cite{bayliss_ngts-1b_2018}. We used the Python-based tool \texttt{masterfile}\footnote{\href{https://github.com/AntoineDarveau/masterfile}{\nolinkurl{github.com/AntoineDarveau/masterfile}}}, which relies on data from the NASA Exoplanet Archive\footnote{\href{https://exoplanetarchive.ipac.caltech.edu}{\nolinkurl{exoplanetarchive.ipac.caltech.edu}}} to obtain the population of transiting exoplanets with $M_{\rm p} > 0.5$ \Mjup{} and with an incident flux smaller than 50 times that of the Earth, yielding a sample of 21 planets. We then used the distribution of their radii to derive a Gaussian prior with a mean of 1.010 \Rjup{} and a standard deviation of 0.331\,R$_{\rm Jup}$. 

\begin{figure}[htpb!]
    \centering
    \includegraphics[width=0.99\linewidth]{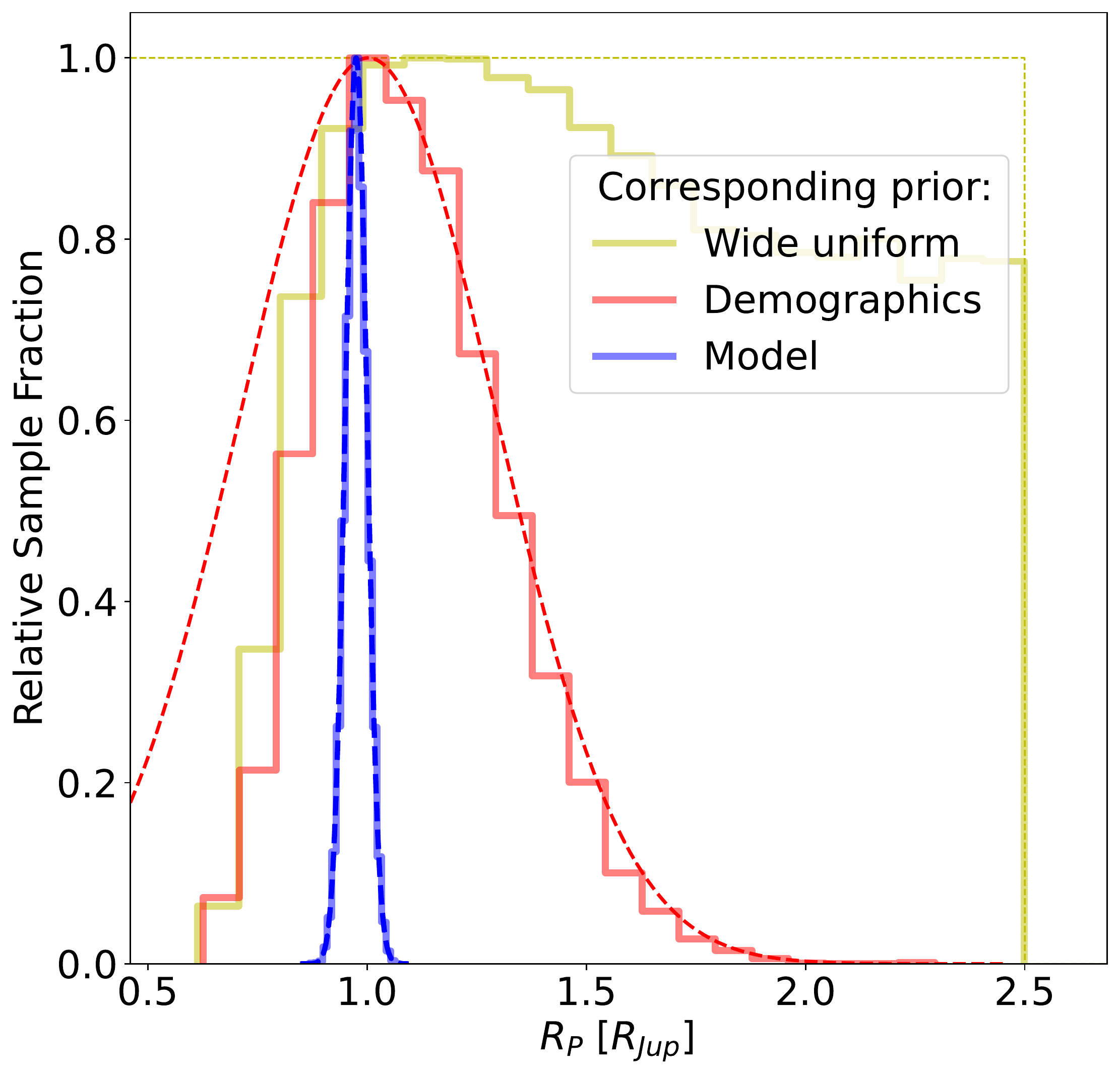}
    \caption{Planetary radius posterior distribution for three different priors. The solid lines show the histogram of the posterior distributions and the corresponding prior probability density functions are shown as dashed lines of the same color. The prior from evolutionary models (blue), as described in Section \ref{subsection:apeculiartemperatureregime}, is $\mathcal{N}(0.975, 0.025)$ and yields $R_{\rm p} = 0.976 \pm 0.025$ \Rjup{}. The prior derived from the confirmed exoplanet population (red) is $\mathcal{N}(1.010, 0.331)$ and yields $R_{\rm p} = 1.09^{+0.24}_{-0.20}$ \Rjup{}. The broad uniform prior $\mathcal{U}(0.0, 2.5)$ (yellow) yields $R_{\rm p} = 1.56^{+0.63}_{-0.53}$ \Rjup{}.}
    \label{fig:rp_hist}
\end{figure}

We sampled the joint posterior distribution of the 15 model parameters, $\{$$P$, $T_0$, $e$, $\omega$, $K$, $R_{\rm p}$, $b$, $M_{\rm s}$, $R_{\rm s}$, $\gamma$, $f_{0, \text{TESS}}$, $f_{0, \text{WBRO}}$, $f_{0, \text{CAO}}$, $u_{1, \text{WBRO}}$, $u_{2, \text{WBRO}}$$\}$, with the No-U-Turn Sampler (\texttt{NUTS}) from \texttt{PyMC3}. We ran four chains with 4000 tuning steps and 4000 draws. The planetary radius posterior distribution for each of the three priors investigated is shown in Figure \ref{fig:rp_hist}. All priors yield a posterior value for $R_{\rm p}$ close to $1~\Rjup{}$, but the width of these distributions varies significantly depending on the strength of the prior. Parameters other than $R_{\rm p}$ and $b$ were not significantly affected by this change of prior. For this reason, in the following, we only report posterior distributions on parameters that used the "Demographics" prior on $R_{\rm p}$. This prior provides an insight on reasonable $R_{\rm p}$ values without completely excluding the possibility of an inflated radius due to irradiation. 

\begin{figure*}[ht!]
    \centering
    \includegraphics[width=\linewidth]{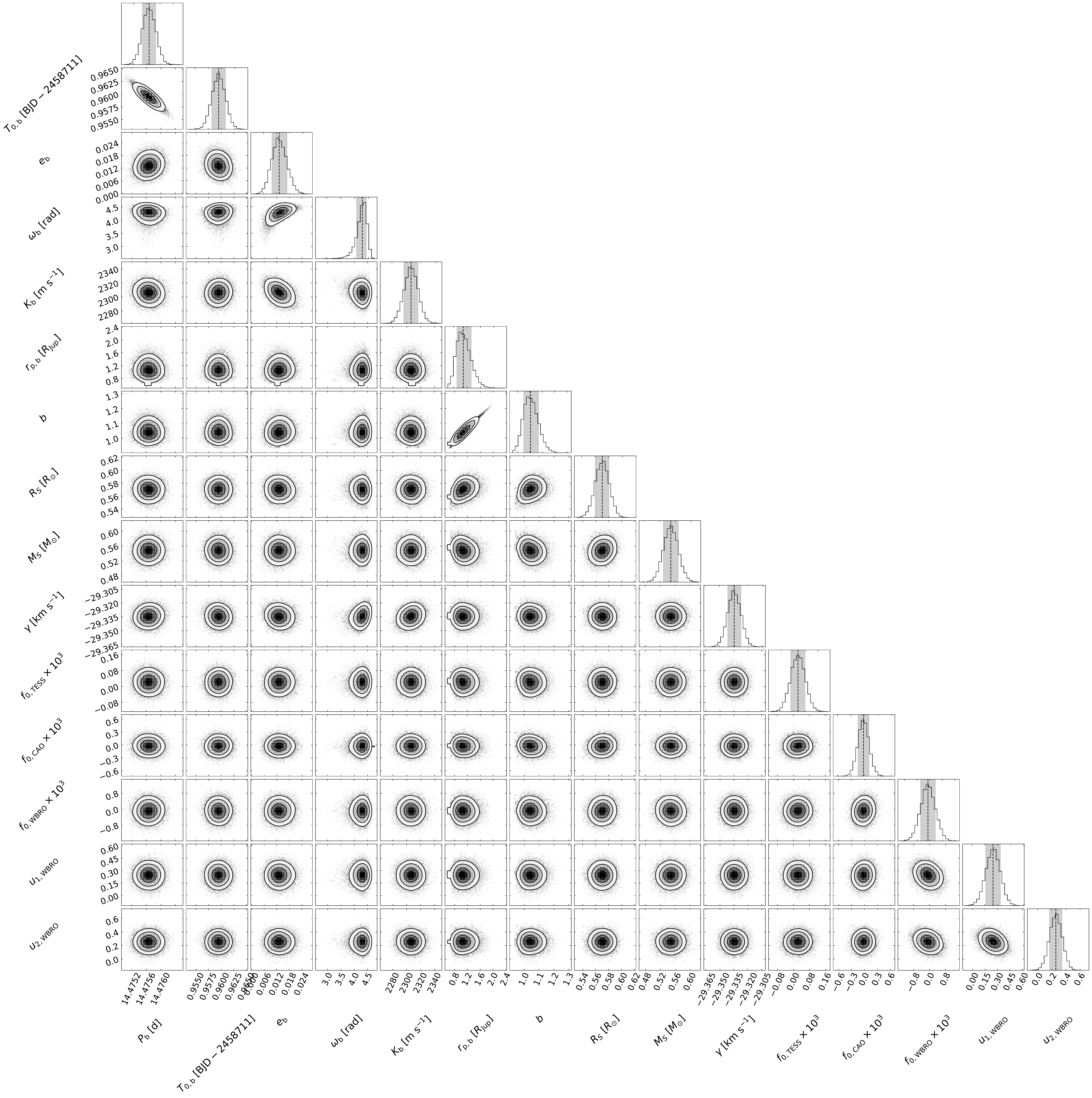}
    \caption{Posterior distribution of the joint RV and transit model parameters. The dashed lines and grey shaded areas show the median and the 1$\sigma$ interval for each parameter.}
    \label{fig:corner_joint}
\end{figure*}

For each parameter, Table \ref{tab:orbit} report the final prior distribution used and the median posterior value, along with uncertainties corresponding to the 16$^{\rm th}$ and 84$^{\rm th}$ percentile. Figure~\ref{fig:lightcurves} shows the resulting transit model on each data set. In Figure~\ref{fig:rv_fit}, the full RV time series and the phase-folded RVs are both shown along with the best-fit model and the residuals. The joint posterior distribution is shown in Figure~\ref{fig:corner_joint}.

% value for the raduis:
% np.sqrt(((3799/5778)**4)/0.0614)

\begin{deluxetable}{ccc}
\tablenum{3}
\tablecaption{Model parameters for the TOI--1278 system\label{tab:orbit}}
\tablehead{
\colhead{Quantity} & \colhead{Prior} & \colhead{ Posterior} 
}
\startdata
        %% Fitted parameters (planet/orbit)
        $P$ [day]                       & $\mathcal{U}(13.476, 15.476)$     & $14.47567 \pm 0.00021$      \\
        $T_0$ [BJD - 2457000]           & $\mathcal{U}(1711, 1713)$         & $1711.9595 \pm 0.0013$      \\
        $R_{\rm p}$ [\Rjup{}]                 & $\mathcal{N}(1.010, 0.331)$      & $1.09^{+0.24}_{-0.20}$ \\
        $b$                             & $\mathcal{U}(0, 1+R_{\rm p}/R_S)$       &$1.04^{+0.06}_{-0.05}$    \\
        $M_{\rm s}$ [\Msun{}]                 & $\mathcal{N}(0.55, 0.02)$         & $0.54 \pm 0.02$           \\
        $R_{\rm s}$ [\Rsun{}]                 & $\mathbf{\mathcal{N}(0.573, 0.012)}$         &  $0.57 \pm 0.01$          \\
        $e$                             & $\mathcal{U}(0, 1)$               & $0.013 \pm 0.004$         \\
        $\omega$ [rad]                  & $\mathcal{U}(0, 2\pi)$            & $4.29^{+0.15}_{-0.21}$    \\
        $K$ [\mps{}]                    & $\text{LogU}(10^{-4}, 10^4)$      & $2306 \pm 10$             \\
        $\phi_{\rm sec}$                & ---                               & $0.496 \pm 0.001$         \\
        $M_{\rm p}$ [\Mjup{}]                 & ---                               & $18.5 \pm 0.5$            \\
        $i$ [deg]                       & ---                               & $88.3 \pm 0.1$            \\
        $a$ [AU]                        & ---                               & $0.095 \pm 0.001$         \\
        $\rho_{\rm p}$ [g~cm$^{-3}$]          & ---                               & $18^{+14}_{-8}$          \\
        $\gamma$ [k\mps{}]              & $\mathcal{U}(-35, -25)$           & $-29.334 \pm 0.007$           \\
        $f_{0,\text{TESS}} \times 10^3$ & $\mathcal{N}(0, 10 000)$          & $0.02 \pm 0.04$           \\
        $f_{0,\text{WBRO}} \times 10^3$ & $\mathcal{N}(0, 10 000)$          & $-0.08 \pm 0.46$            \\
        $f_{0,\text{CAO}} \times 10^3$  & $\mathcal{N}(0, 10 000)$          & $-0.02 \pm 0.12$           \\
        $u_{1,\text{WBRO}}$             & $\mathcal{N}(0.3, 0.1)$           & $0.25 \pm 0.09$            \\
        $u_{2,\text{WBRO}}$             & $\mathcal{N}(0.3, 0.1)$           & $0.26 \pm 0.09$            \\
\hline
\enddata
%\tablecomments{}
\end{deluxetable}

Along with a planetary mass $M_p = 18.5 \pm 0.5$~\Mjup{}, a radius of $1.09^{+0.24}_{-0.20}~\Rjup{}$ yields a density $\rho_p=18^{+14}_{-8}$~g~cm$^{-3}$.
We find a slightly eccentric orbit ($e = 0.013 \pm 0.004$). We also test a circular orbit model with $e=0$ and $\omega=0$ following the same modeling procedure as above. All parameters in this circular model are consistent with the values obtained with the eccentric model. However, the eccentric model is strongly favored by the Bayesian information criterion (BIC), with $\Delta \mathrm{BIC} = 9$. Using the eccentric model, we constrain the time of secondary eclipse to be $T_{\mathrm{sec}} = 1719.143 \pm 0.015$ (BJD - 2\,457\,000). This corresponds to an orbital phase of $0.496 \pm 0.001$. Orbital evolution and the circularization timescale are discussed in Section~\ref{sec:circular}. 

We also performed a sequential analysis where the transits were modeled first, and the resulting $P$ and $T_0$ posterior distributions were used as priors in the RV Keplerian model. This approach has the disadvantage of not accounting for possible impacts of the eccentricity (which is better constrained by the RV data in this case) on the transit model. However, we note that results from both approaches are in excellent agreement, yielding similar posterior distributions with median values consistent within 1$\sigma$. This is notable as the sequential analysis does not rely on the stellar parameters as inputs.

Combining the proper motion and parallax from Gaia~eDR3 with the systemic radial velocity measurement ($\gamma = -29.334 \pm 0.007$\,\kms) derived in this section makes it possible to determine the $UVW$ space velocities of TOI--1278 with an unprecedented precision (the Gaia~eDR3 radial velocity for TOI--1278 is $-29.6 \pm 1.1$\kms). However, doing so requires correcting two systematic effects that become important with meter-per-level instrumental precision: gravitational reddening \citep{1917SPAW.......142E} and convective blueshift (e.g., \citealt{meunier_variability_2017}). Our mass and radius estimations allow us to calculate the expected gravitational redshift of $0.63 \pm 0.03$\,\kms. Estimating the effect of convective blueshift with a high accuracy is much harder, as empirical data on that subject is sparse, especially for M dwarfs. \cite{lohner-bottcher_convective_2019} estimated that this effect blueshifts the perceived radial velocity of the Sun by about $0.3 \pm 0.2$\,\kms. Only one recent study by \cite{baroch_carmenes_2020} investigated this effect for YZ~Cm, a binary M4 dwarf, and found that it ranges from +0.007 to +0.247\,\kms, corresponding to a very slight redshift rather than a blueshift that is expected for most stars. \citep{dai_measuring_2018} investigated this effect using a larger sample from Gaia~DR2, and found an empirical blueshift of $-0.2 \pm 0.2$\,\kms\ for early K dwarfs. We therefore estimate that the convective blueshift for a field-aged M0 star is likely of about $0.0 \pm 0.2$\,\kms, as a middle case between those of YZ~CMi and early K dwarfs. The total redshift to which TOI--1278 is subjected is consequently $0.6 \pm 0.2$\,\kms. We therefore estimate a heliocentric radial velocity of $-29.9 \pm 0.2$\,\kms\ for TOI--1278, and we use this value to calculate the space velocities listed in Table~\ref{tab:parallax}. Overall, the space velocities of TOI--1278 are located just within the 2$\sigma$ velocity dispersion of the Galacic thin disk of \cite{bensby_elemental_2003}, indicating that TOI--1278 is likely not much older than $\approx$\,8\,Gyr.

\section{Discussion}\label{section:discussion}
%\subsection{Parameters of TOI--1278\,B}\label{subsection:parametersoftoi1278}

\subsection{On the Paucity of Close-In BD Companions to M Dwarfs}\label{subsection:paucityofclose-inBDcompanions}

\begin{figure*}[!htbp]
    \centering
    \includegraphics[width=0.999\linewidth]{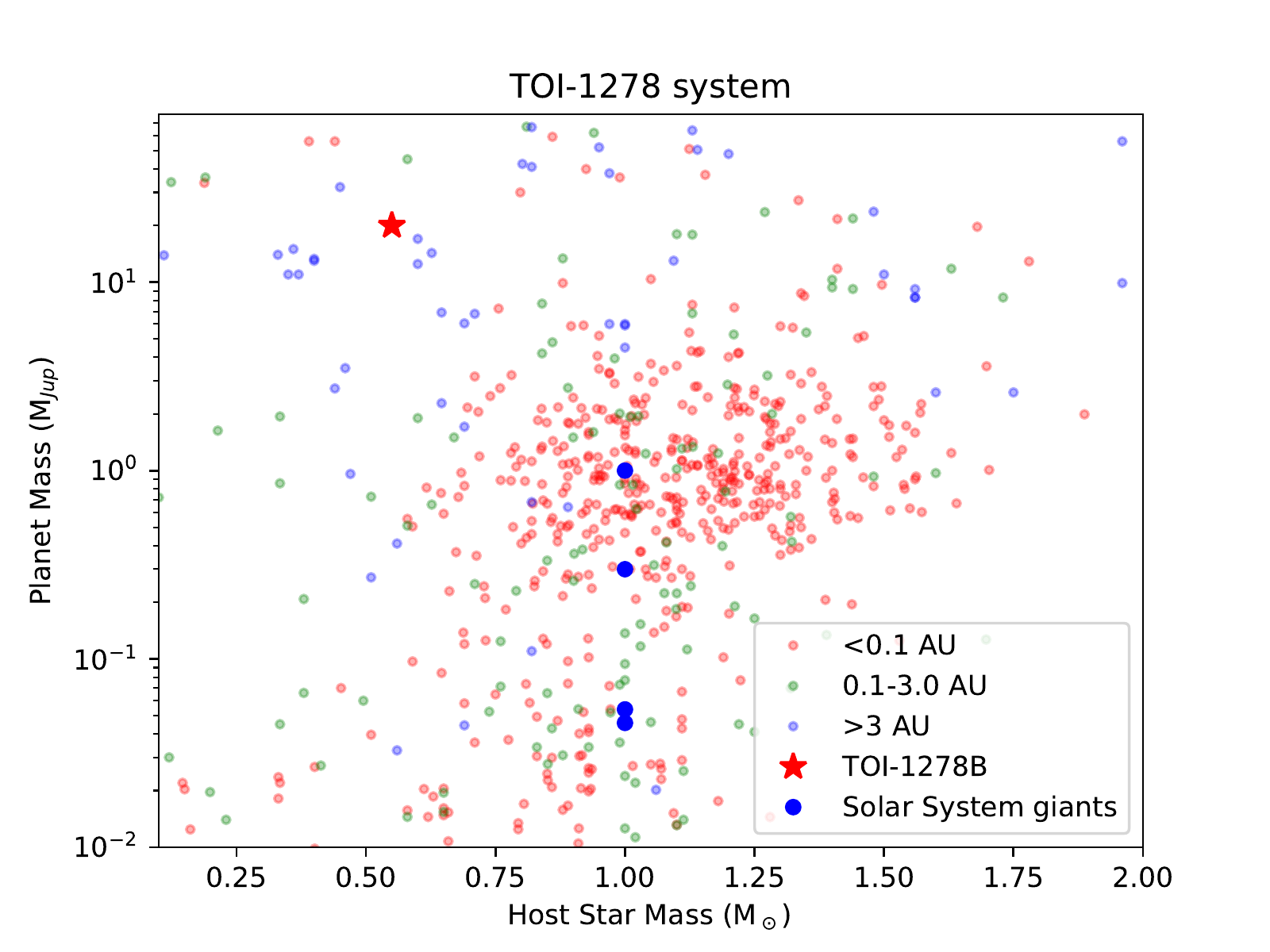}
    \caption{Substellar companion mass as a function of stellar host mass. Close-in companions (red dots) around $<$0.5\,M$_\odot$ stars tend to be lighter than 0.1\,\Mjup. There are also a handful of massive ($>$40\,\Mjup) brown dwarfs in close-in orbits to M dwarfs. (Data from \url{http://exoplanet.eu}) }
    \label{fig:mass_mass}
\end{figure*}

The number of known brown dwarf companions in close-in orbit around main-sequence stars is relatively small (see full compilation by \citet{mireles_toi_2020}, table~4) although they are easier to find in general than planetary companions. This is expected from a formation point of view, with close-in binaries having mass ratios tending toward unity \citep{bate_predicting_2000}. While most such companions orbit Sun-like stars, TOI--1278 combines relatively rare properties; there are  few close-in ($<$0.1\,AU) companions to M dwarfs that are more massive than Saturn (see Figure~\ref{fig:mass_mass} and Table~\ref{tab:massivecompanions} for a compilation of  0.5-78\,\Mjup\ companions to M dwarfs).

%An order-of-magnitude upper limit on the abundance of TOI-1278-like systems +can be %determined+ from the fact…

An order-of-magnitude upper limit on the abundance of TOI--1278-like systems can be determined from the fact that RV surveys of the solar neighborhood have failed to uncovered similar systems, even though it would have been well above their detection thresholds. The HARPS survey of southern M dwarfs \citep{bonfils_harps_2013} surveyed 102 stars with the most massive planet in the sample (Gl876\,b; \citet{delfosse_closest_1998}) has a mass of 2.64\Mjup, $\sim$7 times lighter than TOI--1278\,B. The CARMENES search for exoplanets around M dwarfs \citep{reiners_carmenes_2018} surveyed 324 nearby stars (with some overlap between samples) and did not report any such massive companion. Overall, one can safely say that the occurrence of such systems is well below 1\,\%.

NGTS-1\,b is the most massive transiting planetary companion to an M dwarf \citep{bayliss_ngts-1b_2018}. It is $\sim20$ times less massive (0.8\,\Mjup) and orbits on a shorter 2.647\,d period. There are a number of comparable (10-15\,\Mjup) companions to M dwarfs on  distant orbits uncovered through direct imaging (e.g., GU Psc\,b, 2M\,J2126-81\,b, CHXR\,73\,b, DH\,Tau\,b, 2M\,0103-55(AB)b; \citealt{naud_discovery_2014, luhman_discovery_2006,deacon_nearby_2016,itoh_young_2005, delorme_direct-imaging_2013}). These companions form well outside the extent of protoplanetary disks around M dwarfs, and the statistical properties in terms of orbital separation and mass  suggests a different formation mechanism than for close-in planetary companions \citep{baron_constraints_2019}. The mass ratio of the TOI--1278 system ($\sim31$) cannot be explained through a process involving a planetary disk; the mass involved is larger than the total gas mass in planetary disks, which has a plateau around 10\,\Mjup \citep{bergin_determination_2018}. As M dwarfs luminosity falls rapidly with decreasing mass ($L_\star\propto M_\star^3$), uncovering companions with  small $q$ values is challenging due to the contrast ratio involved and there are significant biases against the discovery of low-$q$ binaries. Detecting TOI--1278-like systems requires RV follow-ups, as Gyr-old $1$\,\Mjup to $0.1$\,\Msun companions all have radii in the $0.8-1.0$\,R$_{\rm Jup}$; transit alone does not set a significant constraint on the nature of the companion. Conversely, the radial velocity signal of such a companion will be huge, from 100\,m/s to many km/s, implying that the companions to faint host stars can be monitored. The paucity of $10-30$\,\Mjup may also be arising from a bias in the follow-up strategy by community members; with an impact parameter slightly lower, TOI--1278\,B's transit would have a relative depth of $\sim4$\,\% and flagged as a likely eclipsing binary.  A dedicated search for similar objects within the TESS data set complemented with a follow-up of a statistically representative set of objects will be needed to assess the abundance of objects similar to TOI--1278\,B. Current follow-ups, focusing on smaller-radius planets around relatively bright and nearby M dwarfs and avoiding possible eclipsing binaries, are naturally biased against the discovery of objects similar to TOI--1278\,B. The true abundance of such systems therefore remains to be established.

Among transiting companions to M dwarfs, only a handful are in the brown dwarf mass regime. The LHS\,6343 system is composed of a visual binary  of two mid-Ms, (0$\farcs$55 apart; 0.358 and 0.292\,M$_\odot$), one component being itself orbited by a close-in brown dwarf \citep{johnson_lhs_2011}. LHS\,6343\,C orbits on a 12.71\,day orbit around LHS\,6343\,A, and has a kinematically derived mass of $62.7\pm2.4$\,\Mjup. It has a mass ratio of only $\sim$6 with its host star. The  LP 261-75\,b system \citep{irwin_four_2018} consists of a much tighter orbit ($\sim$1.8-day orbit) by a brown dwarf close to the stellar/substellar limit at 65\,\Mjup, and a mass ratio of $\sim$4.6.

\begin{deluxetable*}{l|ccc p{1.2in} p{3in}}
\tablenum{4}
\tablecaption{Known transiting 0.5-78\Mjup companions to M dwarfs with P $<$ 100 days.\label{tab:massivecompanions}}
\tablehead{
\colhead{Host}& \colhead{Mass} & \colhead{SpT} & \colhead{Period}& \colhead{References} & \colhead{Comment } \\
\colhead{}& \colhead{(\Mjup)} & \colhead{} & \colhead{(days)}& \colhead{} & \colhead{ } 
}
\startdata
%HD 41004\,B & $<$25\Mjup& M2 & 1.3 & \citet{santos_coralie_2002} & Orbiting the M dwarf component of a K+M system\\
NGTS-1b &$0.81\pm0.07$&M$0.5\pm0.5$&2.65&\citealt{bayliss_ngts-1b_2018}&Grazing transit, kinematically old host.\\
GJ\,876c & $0.83\pm0.03$& M$4$& 30.23 & \citealt{marcy_pair_2001}&Non-transiting object, absolute mass estimated from dynamics interaction by \citealt{correia_harps_2010}   \\
%HAT-P-23 & 1.34 & K7\tnc & 1.21 & &\\ 
GJ\,876b & $2.64\pm0.04$&M$4$&61.93&\citealt{delfosse_closest_1998}&Non-transiting object \\
TOI-519 & $<14$ &M$3.5^{+1.0}_{-0.5}$ & 1.26 & \citealt{parviainen_toi-519_2020} &Upper limit on mass from light curve analysis. 1.07\,\Rjup consistent with an object as light as 0.5\,\Mjup.  \\
TOI--1278    & $18.4\pm0.5$ & M0 & 14.48 & This work &$ $\\
HD\,41004Bb &  $18.4\pm0.22$ & M2 & 1.32 & \citealt{santos_coralie_2002} & Orbiting the M component of a K+M stellar binary. Only $M \sin{i}$ is known \\
NLTT\,41135 &  $33.7\pm2.8$          & M$5.1\pm0.5$   &  2.89    &\citealt{irwin_nltt_2010}& Transiting one of the components of a 2\farcs3 binary.\\
AD 3116 & $54.2\pm4.3$ & M3.9 & 1.98 & \citealt{gillen_new_2017} & High-probability Praesepe member.\\
RIK\,72     & $59.2\pm6.8$            &  M2.5   &  $\simeq97.8$    &\citealt{david_age_2019} &Member of Upper Scorpius (5-7\,Myr); highly inflated.\\
LHS\,6343\,A & $62.1\pm1.2$ & M\tnb& 12.71& \citealt{johnson_lhs_2011, montet_characterizing_2015} & Transiting one of the components of a 0\farcs55 binary.\\
LP\,261-75  &   $67.6\pm2.1$         & M4.5     &  1.88    & \citealt{irwin_four_2018}&10\% deep transit at low impact parameter.\\
NGTS-7A     &    $75^{+3.0}_{-13.7}$\tna        & M3     &    0.68  &\citealt{jackman_ngts-7ab_2019}& Young (55\,Myr), active, tidally locked in a decaying orbit (remaining lifetime of $5-10$\,Myr). Eclipsing one of the components of a 1\farcs13 binary.\\\hline
\enddata
\tablecomments{\tna The authors explore two scenarios regarding the distance of the NGTS-7 system, leading to significantly different companion masses. The GAIA EDR3 parallax ($6.33\pm0.07$\,mas) for the host star is consistent with scenario $i$ and rules-out the lighter mass of scenario $ii$. \tnb Spectral subtype not specified, but $3431\pm21$\,K temperature \citep{montet_characterizing_2015} for host star corresponds to M$2.5\pm0.5$ \citep{boyajian_stellar_2012}.
}
\end{deluxetable*}

%\todo{A table of close-in $>0.5$\Mjup companions to M dwarfs?}

%
%\begin{itemize}
%\item Relatively rare mass ratio of about 25. Differs significantly from massive planets around A stars that have a mass ratio not too far from Jupiter vs Sun.
%\item Show a host VS planet mass diagram and TOI--1278
%\item Systems overlooked as a b=0 transit would have a depth of %$\sim4\%$, maybe above the threshold to make it as a TOO?
%\item fetch exoplanets.eu and construct figure with : x=host mass [0.1-2Mjup, log], y=planet mass [1/100,70 mjup, log], have a color code around transiting ones.
%\end{itemize}

\subsection{Eccentricity and circularization timescale\label{sec:circular}}
As discussed in Section~\ref{sec:dataanalysis}, we find that a slightly eccentric orbit ($e=0.013\pm0.004$) is preferred over a circular one. We compute the expected timescale for tidal orbital circularization of TOI--1278\,B by using \hbox{equation (25)} of \citet{goldreich_q_1966} as rewritten by \citet{patra_apparently_2017}:
\begin{equation}
\tau_e = \frac{e}{de/dt} = \frac{2Q_p}{63 \pi} \frac{M_p}{M_\star} \left(\frac{a}{R_p}\right)^5 P_{\rm orb}
\end{equation}

$Q_p$ being the tidal dissipation parameter of the companion, depending on its internal structure. We assume here that $Q_p = 10^{3.5}$ as determined by \citet{heller_tidal_2010} for the brown dwarf. We find $\tau_e$ of a similar comparable to the age of the Universe ($\tau_e \sim 16 \times10^9$ yr), suggesting that the orbit has not had time to fully circularize. This further implies that eccentric orbit does not require the presence of an external exciting companion to be explained. We  note  that $\tau_e$ scales as the fifth power of $R_{\rm p}$, and that this radius is currently only constrained by models for M and brown dwarfs.

\subsection{Model-Derived Properties and Characterization Prospects}
\label{subsection:apeculiartemperatureregime}
Brown dwarfs are ever cooling objects that gradually lose the heat from their initial contraction. In absence of a luminosity measurement, only the combination of age and mass provides some constraint on the BD temperature. While the mass of TOI--1278\,B is  well constrained at $18.5\pm0.5$\Mjup, its age is not. Its $\gtrsim100$\,Myr lower limit is set by the uninflated radius of the parent star, while the thin-disk kinematics sets a limit at $\lesssim$\,9\,Gyr \citep{del_peloso_age_2005}. Considering a formation rate that is uniform through time in the galactic disk, it is unlikely to be at the lower end of this range when the temperature and radius rapidly decrease.  Assuming a uniform prior distribution in age for the galactic disk, the corresponding  1-$\sigma$ confidence level interval for the age of TOI--1278 is therefore $1.4-7.6$\,Gyr (14$^{\rm th}$ and 86$^{\rm th}$ percentiles of the distribution).

We used the latest evolutionary models for cool T and Y dwarfs \citep{phillips_new_2020} to determine the range of properties expected for TOI--1278\,B. These constraints on the age of the system point toward a cool-down temperature ranging from 600\,K to 380\,K (1-$\sigma$ confidence), but as high as 1360\,K for the lower end of the age range; these values correspond to a bulk luminosity ranging from $10^{-4.4}$ to $10^{-6.8}$\,L$_\odot$, to be compared to the host star luminosity of $5.9\times10^{-2}$\,L$_\odot$. Expressed in terms of brown dwarf spectral sequence, TOI--1278\,B is therefore expected to have a spectral type of L7--L8  if we assume an age of 100\,Myr, T8--T9 for an age of 1.5\,Gyr or Y1 for an age of $7.6$\,Gyr \citep{faherty_population_2016, kirkpatrick_preliminary_2019}.

%https://ui.adsabs.harvard.edu/abs/2016ApJS..225...10F/abstract
% p = [1.546e-4,-1.606e-2,6.318e-1,-1.191e1,1.155e2,-7.005e2,4.747e3]
%L0 2248.6000000000004
%L1 2102.3780706
%L2 1958.9160064000016
%L3 1821.8288914
%L4 1694.8016255999992
%L5 1581.0531250000008
%L6 1482.9118335999992
%L7 1401.5025474000036
%L8 1336.5445504000018
%L9 1286.2610626000023
%T0 1247.4000000000042
%T1 1215.3660466000006
%T2 1184.4640384000013
%T3 1148.253659399997
%T4 1100.0154495999896
%T5 1033.3281250000086
%T6 942.7572095999758
%T7 824.6549794000025
%T8 678.0717184
%T9 505.7782865999643

Assuming the $1~\sigma$ confidence interval for the age, the radius of TOI--1278\,B is remarkably well constrained at $0.975\pm0.025$\,R$_{\rm Jup}$; degeneracy pressure dominates the radius evolution of objects in this age and mass range, implying that the object's radius is (nearly) constant through time. This provides an important prior for transit light curve fitting; the $b\sim1$ impact parameter implies that the light curve alone sets  little constraint on the planetary radius, with a strong degeneracy between $b$ an $R_{\rm p}$ as seen from the diagonal in the corresponding panel of Figure~\ref{fig:corner_joint}. A much larger planet (up to $R_{\rm p}\sim2$\,\Rjup) with an impact parameter of $b\sim1.2$ would be allowed from transit light curve analysis alone, but is rejected by evolutionary models.

One important output of evolutionary models is the per-band flux, which can be combined with the host star photometry (see Table~\ref{tab:model}) to derive a flux ratio. While the system is far too close to be directly resolved with instrumentation in the foreseeable future ($\sim1$\,mas at elongation), it is more likely to be detected through cross-correlation techniques or eclipse photometry or spectroscopy. In the most optimistic case, the contrast ratio would be at the part-per-thousand level in $K$ band and, most likely, significantly more challenging (See Table~\ref{tab:model}). Overall, the prospects for eclipse spectroscopy of TOI--1278\,B are poor.

One can estimate the scale height of the atmosphere of TOI-1278\,B and the corresponding transit-spectroscopy signal. As its atmosphere is most-likely hydrogen-dominated, one can safely assume a mean-molecular weight of 2 AMU. Assuming a temperature of $380-1300$\,K, which would correspond to the  plausible age range, the scale height of the atmosphere would be $5-17$\,km, making it an impractical target for transit spectroscopy.

The expected Rossiter-McLaughlin (RM) effect \citep{rossiter_detection_1924,mclaughlin_results_1924} for this system can be estimated. Assuming a \vsini\, = 1.1 $\pm$ 0.9\,km/s and a transit depth of $\sim1\%$, the maximum possible signal for the effect would be at the $7\pm6$\,m/s level. This is below the sensitivity of our SPIRou observations and would be challenging on such a faint target with current facilities. Considering that the transit is grazing, the  signal is likely to be much smaller if the spin-orbit angle is close to 0$^\circ$. A handful of hot Jupiters in polar orbits are known (e.g., \citealt{addison_nearly_2013}), but these orbit stars hotter than the Sun. Stars such as TOI--1278 have thicker convective layers that dampen orbital obliquity, suggesting that it is unlikely that this system displays has a strong inclination and that it's RM effect can be measured considering its transit configuration \citep{winn_hot_2010}.

\begin{figure*}[!htbp]
    \centering
    \includegraphics[width=0.95\linewidth]{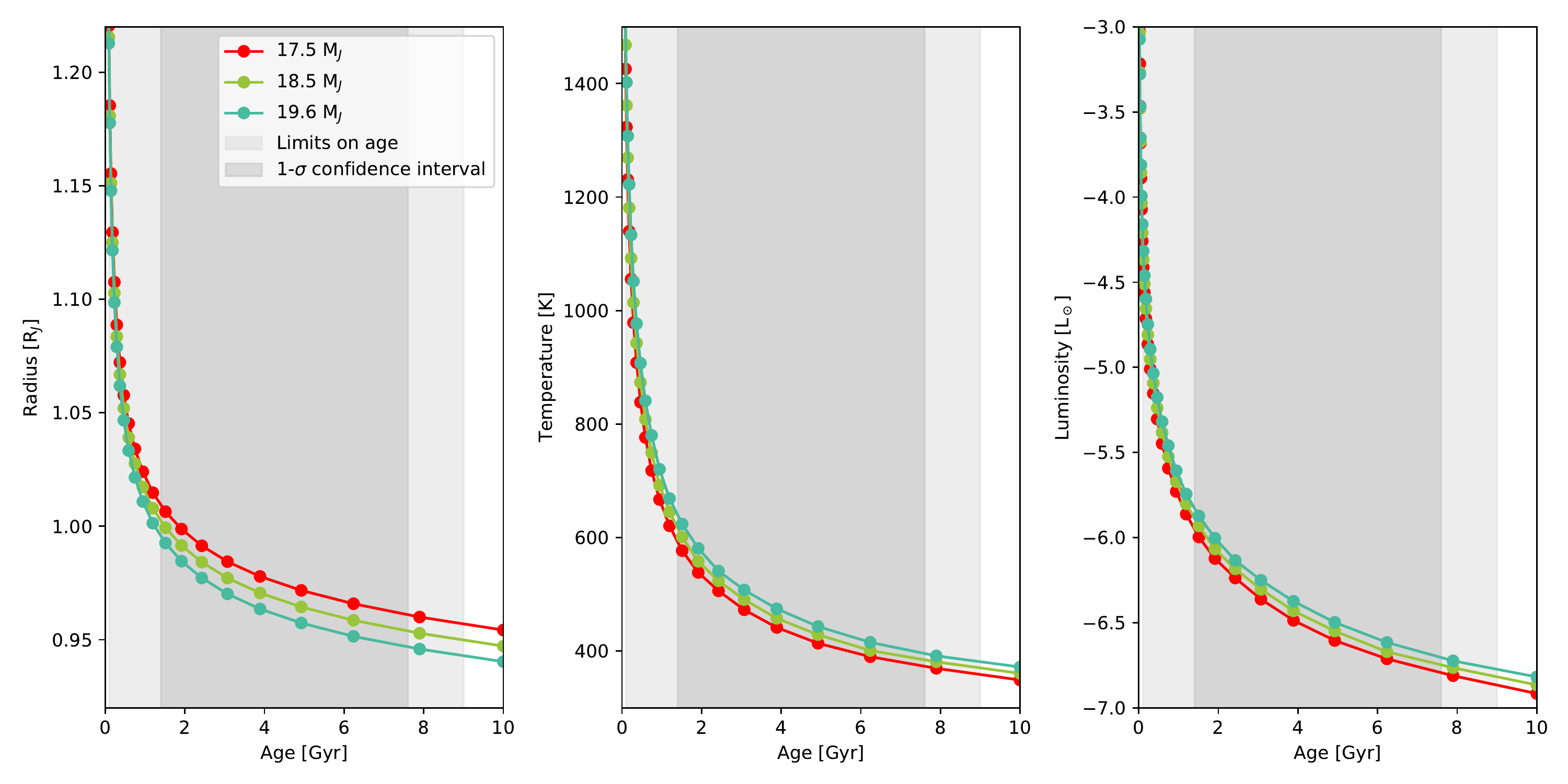}
    \caption{Radius, temperature and luminosity of TOI--1278\,B as a function of age and mass. The derived values are only mildly dependent on the mass, as it is  well constrained from the PRV measurements. Unless the system is at the lowest allowed age value, somewhat unlikely but not impossible, the radius, temperature and luminosity are constrained to within $\pm2.5$\%, $\pm$110\,K and $\pm$0.4\,dex respectively.}
    \label{fig:radius_age}
\end{figure*}

\begin{deluxetable}{l|ccc}
\tablenum{5}
\tablecaption{Model properties of TOI--1278\,B at different ages for a mass of 18.5\,\Mjup. These values are interpolated from \citet{phillips_new_2020} brown dwarf models.\label{tab:model}}
\tablehead{
\colhead{} & \colhead{ {0.1\,Gyr}} & \colhead{1.5\,Gyr} & \colhead{ 7.9\,Gyr } 
}
\startdata
Temperature & 1362 K & 601 K & 381 K\\ 
Luminosity & -4.37 L$_\odot$ & -5.93 L$_\odot$ & -6.76 L$_\odot$\\ 
Gravity & 4.55  & 4.69 & 4.74 \\ 
Radius & 0.118 R$_\odot$  & 0.100 R$_\odot$ & 0.095 R$_\odot$ \\ 
$\Delta$$J_{MKO}$ & 6.878 & 10.992  & 15.285  \\ 
$\Delta$$H_{MKO}$ & 7.330 & 12.551  & 16.747  \\ 
$\Delta$$K_{MKO}$ & 7.640 & 13.113  & 18.720  \\ 
$\Delta$$W1$ & 7.520 & 11.768  & 14.791  \\ 
$\Delta$$W2$ & 6.028 & 8.275  & 9.953  \\ 
$\Delta$$W3$ & 5.134 & 7.620  & 9.157  \\ 
\hline
\enddata
%\tablecomments{}
\end{deluxetable}

%\begin{itemize}
%\item TOI--1278\,B has an equilibrium temperature of X, cooling tracks put its cool-down temperature at about Y. Possibility of testing giant planet atmosphere models in a regime that differs from hot Jupiters
%\end{itemize}

%\subsection{Characterization Prospects}\label{subsection:characterizationprospects}
%\todo{
%\begin{itemize}
%\item expected reflected light contrast and visibility as an SB2?
%\item expected contrast in various bands, use new 2020 BD models; table with contrasts.
%\item Eclipse with MIRI, predict contrast with new models at $\sim$10\,$\mu$m. Check for NIRISS contrast, maybe doomed ?
%\item How good is our knowledge of orbital parameters and do we constrain the orbital fit well enough to predict eclipse for MIRI.
%\end{itemize}
%}

We note that TOI--1278\,B is expected to be among the astrometric detections of substellar companions in Gaia. The semi-major axis of 0.1\,AU and 75\,pc distance implies that the orbit subtends a 1.3\,mas angle and that astrometric orbital motion of the host star is 50\,$\mu$as. At $G\sim12.7$, the per-epoch astrometric accuracy is at the {34.2}\,$\mu$as level \citep{perryman_astrometric_2014}. The presence of the companion most likely explains the \texttt{astrometric\_excess\_noise\_sig} = 9.09 value in the EDR3 catalog. This dimensionless parameter measures the significance of the excess in noise relative to the astrometric uncertainties when fitting a simple parallax models. For good fits of single stars, one would expect half of the targets to have a value of zero as the residuals to the fit will be (slightly) smaller than expected from uncertainties. The remaining targets will have values displaying a cumulative distribution corresponding to a Gaussian with a width of unity. Sources with a value above 2 are considered as likely having a significant noise excess in their parallax solutions. While Gaia does not have the proper time-sampling to probe short periods such as that of TOI--1278, the known orbital period from RV measurements could guide a determination of the actual astrometric orbit.

\clearpage
\acknowledgments
We dedicate this paper to the memory of France Allard, who until her untimely passing did so much to advance the modeling of brown dwarf atmospheres.\\
\\
Based on observations obtained at the Canada-France-Hawaii Telescope (CFHT) which is operated from the summit of Maunakea by the National Research Council of Canada, the Institut National des Sciences de l'Univers of the Centre National de la Recherche Scientifique of France, and the University of Hawaii. The observations at the Canada-France-Hawaii Telescope were performed with care and respect from the summit of Maunakea which is a significant cultural and historic site.\\
\\
This work has been carried out within the framework of the National Centre of Competence in Research PlanetS supported by the Swiss National Science Foundation.\\
\\
JFD acknowledges funding from the European Research Council under the H2020 research \& innovation programme (grant \#740651 NewWorlds).\\
\\
This work was supported by FCT - Funda\c c\~ao para a Ci\^encia e a Tecnologia through national funds and by FEDER through COMPETE2020 - Programa Operacional Competitividade e Internacionaliza\c c\~ao by these grants: UID/FIS/04434/2019; UIDB/04434/2020; UIDP/04434/2020; PTDC/FIS-AST/32113/2017 \& POCI-01-0145-FEDER-032113; PTDC/FIS-AST/28953/2017 \& POCI-01-0145-FEDER-028953.\\
\\
J.H.C.M. is supported in the form of a work contract funded by Funda\c c\~ao para a Ci\^encia e Tecnologia (FCT) with the reference DL 57/2016/CP1364/CT0007; and also supported from FCT through national funds and by FEDER-Fundo Europeu de Desenvolvimento Regional through COMPETE2020-Programa Operacional Competitividade e Internacionaliza\c c\~ao for these grants UIDB/04434/2020 \& UIDP/04434/2020, PTDC/FIS-AST/32113/2017 \& POCI-01-0145-FEDER-032113, PTDC/FIS-AST/28953/2017 \& POCI-01-0145-FEDER-028953, PTDC/FIS-AST/29942/2017.

XD, TF and I.B.  received funding from the French Programme National de Physique Stellaire (PNPS) and the Programme National de Plan\'etologie (PNP) of CNRS (INSU)

XD, TF and IB acknowledge funding from the French National Research Agency (ANR) under contract number ANR-18-CE31-0019 (SPlaSH).

This work has made use of data from the European Space Agency (ESA) mission {\it Gaia} (\url{https://www.cosmos.esa.int/gaia}), processed by the {\it Gaia} Data Processing and Analysis Consortium (DPAC, \url{https://www.cosmos.esa.int/web/gaia/dpac/consortium}). Funding for the DPAC has been provided by national institutions, in particular the institutions participating in the {\it Gaia} Multilateral Agreement.

Funding for the TESS mission is provided by NASA's Science Mission directorate. We acknowledge the use of public TESS Alert data from pipelines at the TESS Science Office and at the TESS Science Processing Operations Center. This research has made use of the Exoplanet Follow-up Observation Program website, which is operated by the California Institute of Technology, under contract with the National Aeronautics and Space Administration under the Exoplanet Exploration Program. This paper includes data collected by the TESS mission that are publicly available from the Mikulski Archive for Space Telescopes (MAST).

%This research made use of pandas \citep{McKinney_2010, McKinney_2011}

%This research made use of Astropy, a community-developed core Python
%package for Astronomy (Astropy Collaboration, 2018).

This research has made use of the SIMBAD database, operated at CDS, Strasbourg, France. This research has made use of NASA's Astrophysics Data System. This research made use of matplotlib, a Python library for publication quality graphics \citep{hunter_matplotlib_2007}. This research made use of SciPy \citep{scipy_10_contributors_scipy_2020}. This work made use of the IPython package \citep{perez_ipython_2007}. This research made use of Astropy, a community-developed core Python package for Astronomy \citep{the_astropy_collaboration_astropy_2013,the_astropy_collaboration_astropy_2018}. This research made use of NumPy \citep{harris_array_2020}. This research made use of TOPCAT, an interactive graphical viewer and editor for tabular data \citep{taylor_topcat_2005}. This research made use of Astroquery \citep{ginsburg_astroquery_2019}. This research made use of \texttt{ds9}, a tool for data visualization supported by the Chandra X-ray Science Center (CXC) and the High Energy Astrophysics Science Archive Center (HEASARC) with support from the JWST Mission office at the Space Telescope Science Institute for 3D visualization.
\facilities{CFHT(SPIRou)}
\software{astropy \citep{price-whelan_astropy_2018}}

%\section{Appendix information}
\clearpage
\appendix
\section{Table of SPIRou RV Measurements}

\begin{table}[ht!]
    \caption{SPIRou RV measurements binned per epoch}\label{tab:spirou_rv}
    \begin{center}
    \begin{tabular}{ccc}
        \hline\hline
        Time & RV & $\sigma_{\mathrm{RV}}$ \\
        BJD  & [\mps{}] & [\mps{}]\\
        \hline
        2459000.1000 & -28111.4 & 17.2 \\
        2459003.0779 & -30822.7 & 28.5 \\
        2459005.0466 & -31675.7 & 15.6 \\
        2459006.0240 & -31491.6 & 14.9 \\
        2459008.0282 & -29992.7 & 13.3 \\
        2459008.9816 & -29051.9 & 18.3 \\
        2459010.0464 & -28069.4 & 21.4 \\
        2459011.1114 & -27299.0 & 21.5 \\
        2459153.7576 & -29025.5 & 17.7 \\
        2459154.7451 & -28075.8 & 15.9 \\
        \hline
        \end{tabular}
        \end{center}
\end{table}

\clearpage

\bibliography{TOI1278.bbl}{}
\bibliographystyle{aasjournal.bst}

\end{document}